\documentclass[%
 reprint,
 superscriptaddress,
 amsmath,amssymb,
 aps,
prb,
]{revtex4-1}

\usepackage{graphicx}% Include figure files
\usepackage{dcolumn}% Align table columns on decimal point
\usepackage{bm}% bold math
\usepackage{xcolor}
\usepackage{algorithm}
\usepackage{algpseudocode}
\usepackage{url} 
\usepackage{makecell}
\usepackage{multirow}
\usepackage{booktabs}
\usepackage{graphicx}
\usepackage{subfig}
\def\br{{\bm r}}

\def\phib{\phi}
\def\vi{v}
\def\vii{{\bm v}}
\def\Phib{\Phi}
\def\AAb{{\bm A}}
\def\BBb{{\bm B}}
\def\ace{\varphi}
\def\cAA{\tilde{c}}
\def\cBB{c} 
\def\calF{\mathcal{F}}

\def\ord{\nu}
\def\maxord{\nu_{\rm max}}

\newcommand {\pder}[2]{\frac{\partial  #1}{\partial #2} }

\begin{document}

%\preprint{APS/123-QED}

\title{Performant implementation of the atomic cluster expansion (PACE): 
Application to copper and silicon}

\author{Yury Lysogorskiy} 
\affiliation{ICAMS, Ruhr-Universit\"at Bochum, Bochum, Germany}
\author{Cas van der Oord}
\affiliation{Engineering Laboratory, University of Cambridge, Cambridge, CB2 1PZ UK}
\author{Anton Bochkarev} 
\affiliation{ICAMS, Ruhr-Universit\"at Bochum, Bochum, Germany}
\author{Sarath Menon}
\affiliation{ICAMS, Ruhr-Universit\"at Bochum, Bochum, Germany}
\author{Matteo Rinaldi} 
\affiliation{ICAMS, Ruhr-Universit\"at Bochum, Bochum, Germany}
\author{Thomas Hammerschmidt} 
\affiliation{ICAMS, Ruhr-Universit\"at Bochum, Bochum, Germany}
\author{Matous Mrovec} 
\affiliation{ICAMS, Ruhr-Universit\"at Bochum, Bochum, Germany}
\author{Aidan Thompson}
\affiliation{Center for Computing Research, Sandia National Laboratories, Albuquerque, New Mexico 87185, USA}
\author{G\'{a}bor Cs\'{a}nyi}
\affiliation{Engineering Laboratory, University of Cambridge, Cambridge, CB2 1PZ UK}
\author{Christoph Ortner}
\affiliation{Department of Mathematics, University of British Columbia, Vancouver, BC, Canada V6T 1Z2}
\author{Ralf Drautz} 
\affiliation{ICAMS, Ruhr-Universit\"at Bochum, Bochum, Germany}

\date{\today}% It is always \today, today,

\begin{abstract}
The atomic cluster expansion is a general polynomial expansion of the atomic energy in multi-atom basis functions. Here we implement the atomic cluster expansion in the performant C++ code \verb+PACE+ that is suitable for use in large scale atomistic simulations. We briefly review the atomic cluster expansion and give detailed expressions for energies and forces as well as efficient algorithms for their evaluation. We demonstrate that the atomic cluster expansion as implemented in \verb+PACE+ shifts a previously established Pareto front for machine learning interatomic potentials towards faster and more accurate calculations. Moreover, general purpose parameterizations are presented for copper and silicon and evaluated in detail. We show that the new Cu and Si potentials significantly improve on the best available potentials for highly accurate large-scale atomistic simulations. 

\end{abstract}

\pacs{Valid PACS appear here}% PACS, the Physics and Astronomy
\maketitle

%\tableofcontents

\section{Introduction}
Atomistic modelling and simulation requires efficient computation of energies and forces. In recent years, machine learning (ML) based interatomic potentials, parameterized to large data sets of reference electronic structure calculations, have provided particularly successful surrogate models of the atomic interaction energy.  The ML models construct representations of atomic structure that are used in various regression algorithms to predict energies and forces. 

The recently developed atomic cluster expansion (ACE) \cite{Drautz19} provides a complete and efficient representation of atomic properties as a function of the local atomic environment in terms of many-body correlation functions. Because of the completeness of the ACE basis\cite{Dusson20} these may be employed directly using linear regression for the computation of energies and forces. 
Furthermore, using simple nonlinear embedding functions, ACE can represent many classical as well as ML interatomic potentials. For example, the widely used family of Embedded Atom Method (EAM)\cite{Daw84} and Finnis-Sinclair (FS) \cite{Finnis84} potentials may be viewed as a lowest order ACE. Other properties, for example the moments of the density of states, may also be represented, and recursion or moments-based potentials like the bond-order potentials \cite{Pettifor00,Drautz11} expanded in the form of an ACE.

Moreover, there are deep connections between ACE and several ML representations and formulations. The only other known {\em complete} parameterizations, the Moment Tensor Potentials (MTP) \cite{Shapeev16} and the ML potential of \citet{Seko2019MLIP}, are both based on a body-ordered invariant polynomial basis and can be exactly represented by ACE by suitable choice of hyperparameters and an explicit linear transformation. In addition, the Spectral Neighbor Analysis Potential (SNAP) \cite{Thompson15}, the atomic Permutation Invariant Potentials (aPIPs) \cite{Oord20} and descriptors such as the symmetry functions of \citet{Behler11} and the Smooth Overlap of Atomic Positions (SOAP) \cite{Bartok13} can be obtained as special cases or minor variations of the ACE formalism; see Refs.~\cite{Drautz19,Drautz20,Dusson20} and the supplementary information (SI) for further details.

\begin{figure}
    \includegraphics[width=0.75\columnwidth]{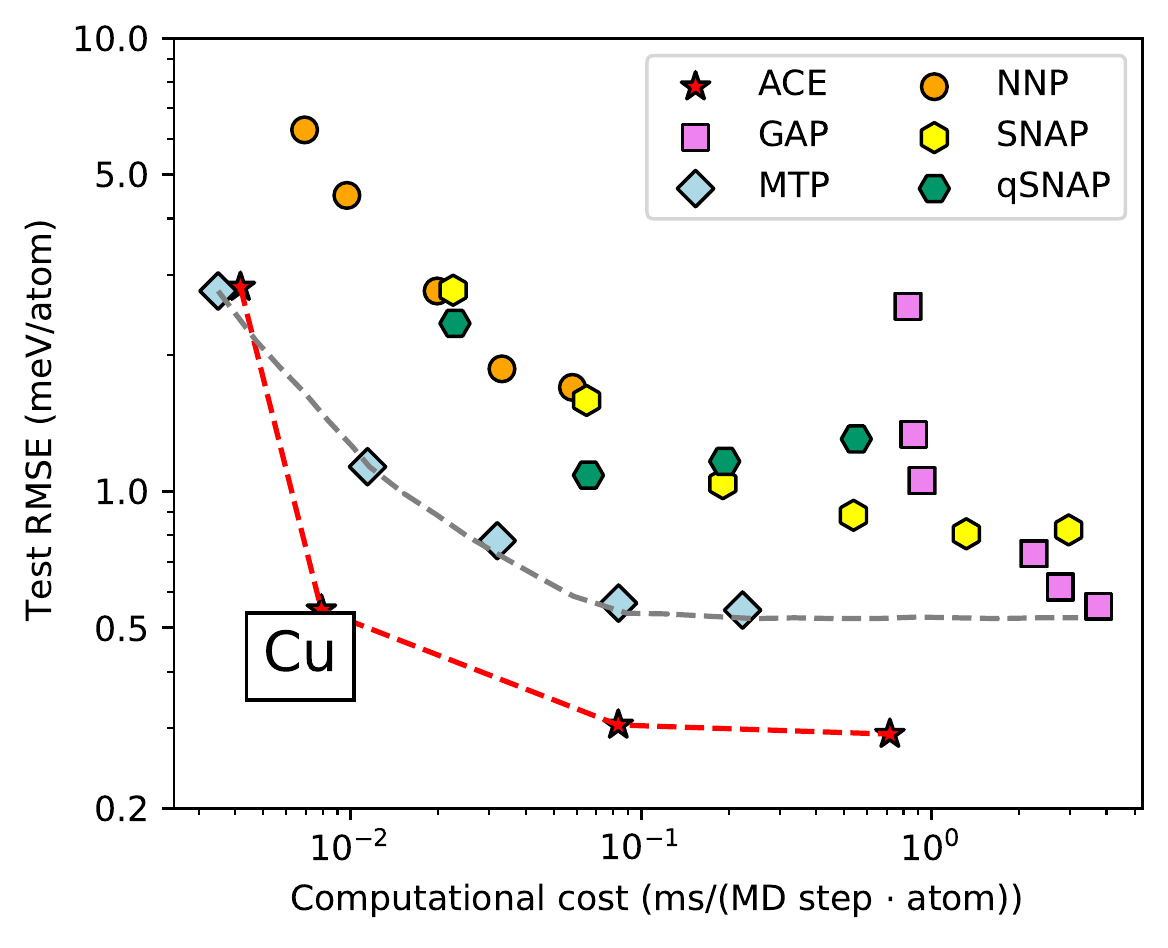}
    \includegraphics[width=0.75\columnwidth]{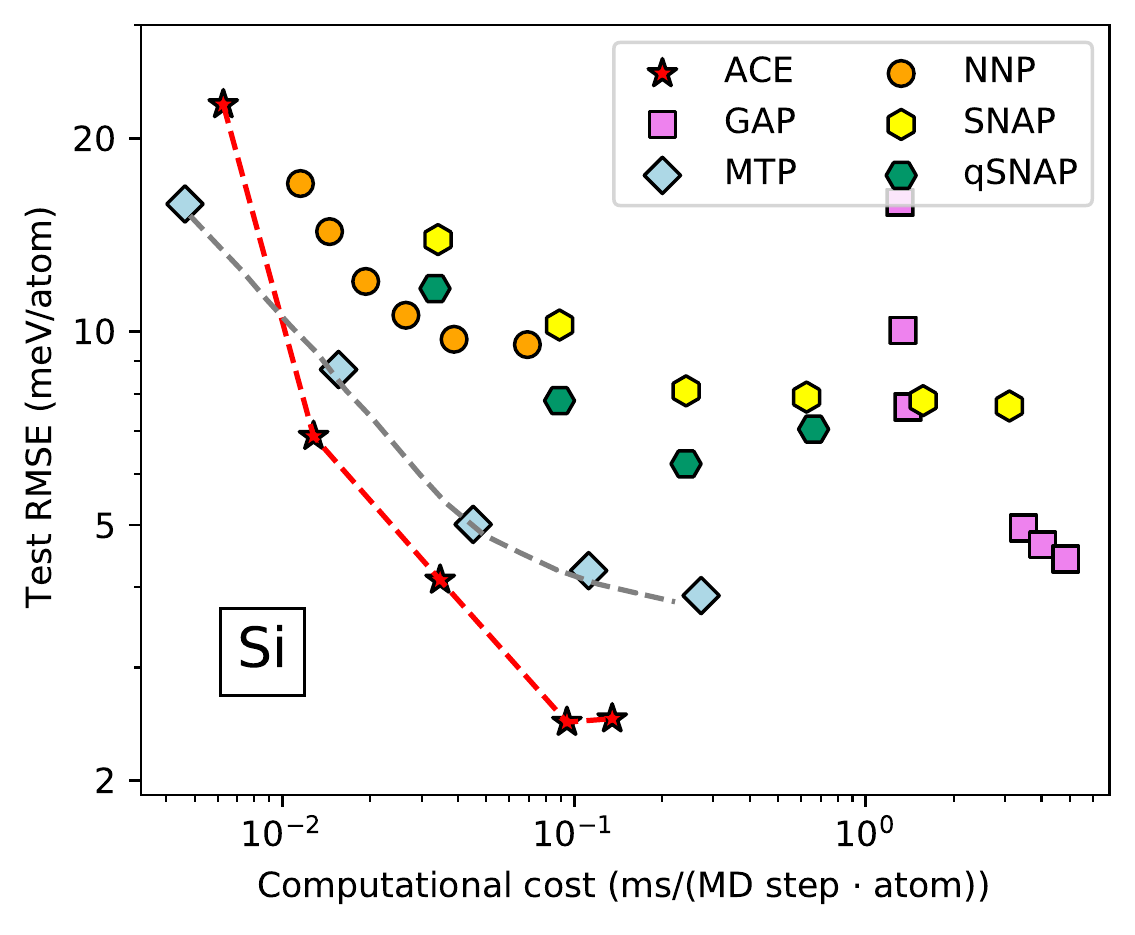}
    \caption{{\bf ACE Pareto front.} Test RMSE versus computational cost for Cu (top) and Si (bottom) for ACE potentials compared to a recent benchmark study~\cite{Zuo20}. The timings from \citet{Zuo20} were reduced by constant factors 0.55 (Cu) and 0.60 (Si) to correct for hardware differences and the new ACE timings then overlaid. 
    }
    \label{fig:Cu_MLEARN_timing}
\end{figure}

Here, we present the {\em Performant Implementation of the Atomic Cluster Expansion} (\verb+PACE+) enabling efficient evaluation of ACE models within the \verb+LAMMPS+ molecular dynamics simulation software package\cite{Plimpton95,LAMMPS}. We demonstrate in Figure~\ref{fig:Cu_MLEARN_timing} for two representative elements, Cu and Si, that \verb+PACE+ lowers the Pareto front of accuracy vs. computational cost that was established for several state-of-the-art ML potentials~\cite{Zuo20}. The details of how these were constructed are provided in the SI. While these benchmarks establish advanced computational performance, we also demonstrate the capacity of the ACE framework to develop highly accurate parameterisations: We present two novel parameterisations of interatomic potentials for Cu and Si that outperform available ML based potentials in terms of performance, accuracy and generalisability.

The fundamental building block from which ACE models are built are {\em atomic properties} $\ace_i$ which are expanded in terms of body-ordered functions from the set of neighbors of each atom $i$,
\begin{equation}
    \ace_i
    = 
    \sum_{\ord = 1}^{\maxord} 
    \sum_{\vii} \cAA_{\vii} \sum_{j_1, \dots, j_\ord} \Phib_{\vii}(\br_{j_1 i}, \dots, \br_{j_\ord i}) \,,
\end{equation}
where $\Phib_\vii$ are $\ord$-order basis functions (each involving coordinates of $\ord$ neighbors) and $\cAA_{\vii}$ the model parameters. It appears as if this incurs an $O(N^\ord)$ computational cost, where $N$ denotes the number of interacting neighbors; however, ACE exploits a much faster evaluation strategy that makes it possible to compute efficiently high body-order terms. This is achieved by (i) projecting the atomic density
\begin{equation}
\rho_i(\br) = \sum_{j \neq i} \delta(\br - {\bm r}_{ji}) \,,
\end{equation}
on atomic basis functions, $\phi_\vi(\br)$, resulting in
\begin{equation}
A_{i\vi} = \sum_{j \neq i} \phi_\vi(\br_{ji}) \,,
\end{equation}
and (ii) choosing a tensor product basis 
\begin{equation}
    \Phib_{\vii}(\br_{1i}, \dots, \br_{\ord i}) 
    = 
    \prod_{t = 1}^\ord \phi_{\vi_t}(\br_{ti}) \,,
\end{equation}
which leads to\cite{Drautz19} 
\begin{equation}
    \sum_{j_1, \dots, j_r} \Phib_{\vii}(\br_{j_1 i}, \dots, \br_{j_r i})
    = 
    \prod_{t = 1}^\ord A_{i\vi_t} \,.
\end{equation}
We call this reformulation the ``density trick'' (also used by \citet{Bartok10} and \citet{Shapeev16} in formulating SOAP and MTP, respectively) and it results in the computational cost of an atomic property $\ace_i$ scaling linearly in $N$ (due to evaluating the $A_{ik}$) and also linearly in $\ord$ (due to evaluating the correlations). Furthermore, in Sec.~\ref{sec:recursiveeval} we present an evaluation scheme that avoids the $\ord$-scaling altogether.

An ACE model may be defined in terms of several atomic properties $\ace_i^{(p)}$, $p = 1, \dots, P$, for each atom $i$. For the simplest linear model of the potential energy one would use just one property, the atomic energy $E_i$, 
\begin{equation}
    E_i = \ace_i^{(1)} \,. \label{eq:Elin}
\end{equation}
A more elaborate model may generalize the pairwise repulsion and the pairwise density of the Finnis-Sinclair potential\cite{Finnis84} to arbitrary many-atom interactions,
\begin{equation}
E_i =  \ace_i^{(1)} - \sqrt{\ace_i^{(2)}}  \,. \label{eq:EFS}
\end{equation}
In general, a large number of different atomic properties that are regarded as descriptors enter a non-linear function
\begin{equation}
\label{eq:generalF}
E_i = \calF(\ace_i^{(1)}, \dots, \ace_i^{(P)})  \,,
\end{equation}
where the non-linearity $\calF$ may be explicit as in the Finnis-Sinclair model, or represent a general approximator such as artificial neural networks, as used by \citet{Behler07}, or a kernel ridge regression model as used in the Gaussian Approximation Potential (GAP) \cite{Bartok10}.

Different non-linearities $\calF$ may be used to incorporate physical or chemical insights in bond formation.
Since the $d$-shell of copper is nearly full, angular contributions are generally small in the bulk, hence copper is modelled well by classical central-force functionals with non-linear EAM or FS type embedding functions that effectively generate high body-order terms\cite{Drautz04, Daw84, Finnis84}. Our parameterization for copper therefore starts from the FS representation of the energy, as in Eq.(\ref{eq:EFS}), but with the two atomic properties not limited to pairwise terms but including many-atom contributions that capture small angular contributions in the bulk and larger angular contributions in small clusters or two-dimensional structures.

On the other hand, the diamond structure of silicon is stabilized by angular contributions over close-packed structures, which highlights the importance of interactions beyond pairwise terms. Many different angle-dependent potentials have been developed for Si.  Perhaps the best known are the Stillinger-Weber potential \cite{Stillinger85} with a linear three-body term and the Tersoff potential \cite{Tersoff88} which includes non-linear functions of three-body contributions. The most accurate potential for silicon to date, the SOAP-GAP model of \citet{Bartok18} is an intrinsically high body-order potential. Here, we present a linear ACE for Si, which may be viewed as a generalization of this potential that includes {\em all} body-order interactions up to some maximum. In this way, ACE is employed in its basic form shown in Eq.~\eqref{eq:Elin}, which simplifies the parameterization considerably and avoids implicit assumptions on the form of non-linear terms that are often present in ML frameworks. 

We carry out a detailed comparison of both our ACE parameterizations to the most reliable models available from literature. For Cu, we compare to the EAM potential by \citet{Mishin2001EAM}, to a recent SNAP \cite{Li2018SNAP} parameterisation as well as the GTINV \cite{Seko2019MLIP} ML potential. For Si, we compare to the GAP that was shown to reproduce a wide range of observable properties for crystalline, liquid and amorphous Si phases\cite{Bartok18}.

\section{Results and Discussion \label{sec:results}}

\subsection{Reference data}

The parameterization for Cu was obtained by matching to the energies and forces of about $50000$ total energy calculations as obtained with density functional theory (DFT) using the PBE\cite{Perdew96} functional as implemented in the \verb+FHI-aims+ code \cite{FHIaims,FHIaims2}. The reference data included small clusters, bulk structures, surfaces and interfaces, point defects and their randomly modified variants. Part of the reference data has been briefly described in Ref.~\onlinecite{Drautz19}, but has been extended significantly for the present parameterization. We employed  \verb+pyiron+\cite{pyiron-paper} for generating part of the reference data.

The parameterization for Si was obtained by fitting to the same extensive silicon database GAP was fit to~\cite{Bartok18}. The database covers a wide range of configurations including crystalline structures, surfaces, vacancies, interstitials and liquid phases. The DFT reference data were generated using the \verb+CASTEP+ \cite{CASTEP} software package. 

\subsection{Parameterization and timing}

We used different parameterization strategies for Cu and Si motivated by their different bond chemistry. In particular, the Si parameterization was obtained from solving a linear system of equations, whereas the Cu fit required non-linear optimization.
The Cu potential has a total number of 2072 parameters, of which 756 are expansion coefficients for each of the two densities, and 560 parameters are used for the radial functions. The DFT reference showed that interactions are smaller than 1 meV when atoms are further than the cutoff distance $r_{c}=7.4$\,\AA\ apart, when rigidly separating slabs.
Parameter optimization led to a fit with an error of 3.2 meV/atom for structures that are within 1 eV of the ground state. This fit was then fine-tuned towards structures close to the ground state, which further decreased the error to 2.9 meV/atom and slightly increased the error of higher energy structures. For Si we used a total of 6827 basis functions parameterized as a linear model, with a maximum body order corresponding to $\nu=4$. These basis functions were selected using the construction outlined in \citet{Dusson20}. We show the silicon ACE matches the accuracy of the general-purpose GAP potential introduced by \citet{Bartok18}. More specifically the energy error for the ACE model was found to be 1.81 meV/atom for structures within 1 eV from the ground state. The corresponding errors for the GAP model are 1.25 meV/atom on the silicon database presented in \citet{Bartok18}. 

To evaluate the computational efficiency of \verb+PACE+ we carried out molecular dynamics (MD) simulations for face-centered cubic (fcc) Cu and diamond Si structures. We found that a single force call takes 0.32 and 0.80~ms/atom, respectively, for the Cu and Si ACE models\footnote{Timings were obtained on a single core of an Intel(R) Xeon(R) Gold 6132 CPU, using the GCC 7.3.0 compiler and LAMMPS version from 4 Feb. 2020.}.
These speeds are sufficiently fast for large scale MD simulations and Monte Carlo sampling, for example, for the computation of phase diagrams. The efficiency of \verb+PACE+ is about two orders of magnitude slower than empirical potentials.

\subsection{Copper}

The ACE for Cu has been comprehensively validated against DFT and available experimental data and compared to three other Cu potentials.  The potentials we chose for the comparison were (i) the EAM potential of Mishin et al.~\cite{Mishin2001EAM}, which exhibits an excellent overall accuracy and is considered as the reference Cu EAM potential, (ii) the SNAP model of Li et al.~\cite{Li2018SNAP}, which was trained to strained crystalline as well as melted Cu phases obtained by ab-initio MD, and (iii) the ML interatomic potential Cu-gtinv-934 (GTINV) of Seko et al.~\cite{Seko2019MLIP}, which was fitted to an extended DFT database of $10^4$ structures and reached RMSE values of 8.2 meV/atom. The EAM and SNAP potentials were computed through the \verb+OpenKIM+ API~\cite{Tadmor2011OpenKIM}.

\begin{figure}
    \includegraphics[width=\columnwidth]{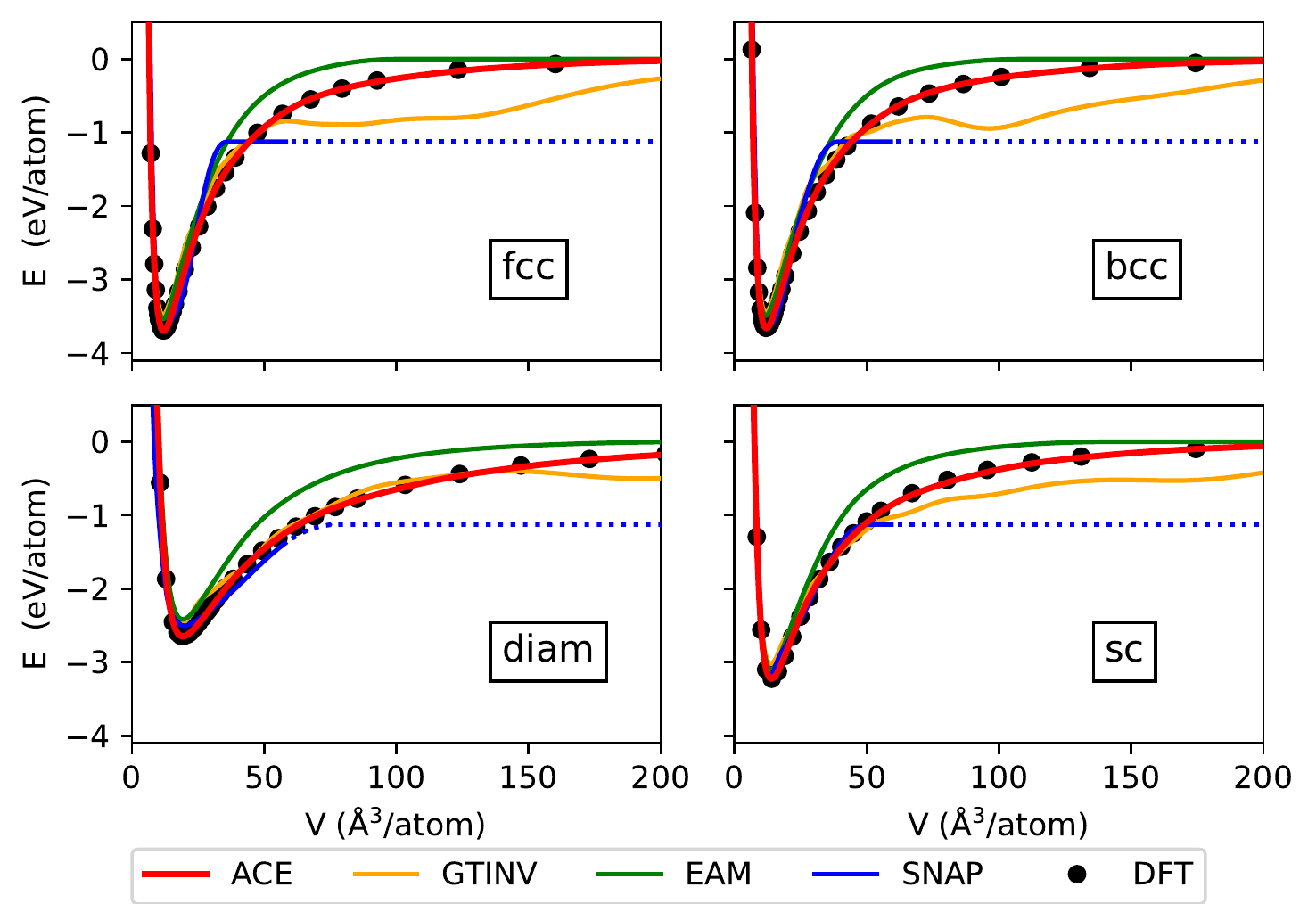}
    \caption{{\bf Cu potentials for large volume changes.} The horizontal axis gives the volume per atom in the fcc, bcc, diamond, and simple-cubic crystal structures. The SNAP cohesive energy was adjusted by a constant shift to match the fcc DFT data. 
    }
    \label{fig:Cu_E_vs_NND}
\end{figure}

We evaluate the models for a broad range of structures and properties that not only exceed beyond the reference data but are also relevant for observable macroscopic behavior of Cu. Fig.~\ref{fig:Cu_E_vs_NND} gives an overview of the binding energy over large volume changes. ACE provides a very good match to the reference data at all distances, while the shorter range of EAM means that interatomic interactions are cut off too early when the atoms are separated. The even shorter cutoff of SNAP leads to abrupt bond breaking, illustrating that the cohesive energy was not fitted in the construction of the potential and therefore one cannot apply the potential, for example, for gas phase condensation simulations. GTINV shows significant oscillations at larger interatomic distances. These observations also apply to the dimer shown in Fig.~\ref{fig:Cu_fcc_surface_dimer_adatom}.

\subsubsection{Bulk properties}

\begin{figure}
    \includegraphics[width=\columnwidth]{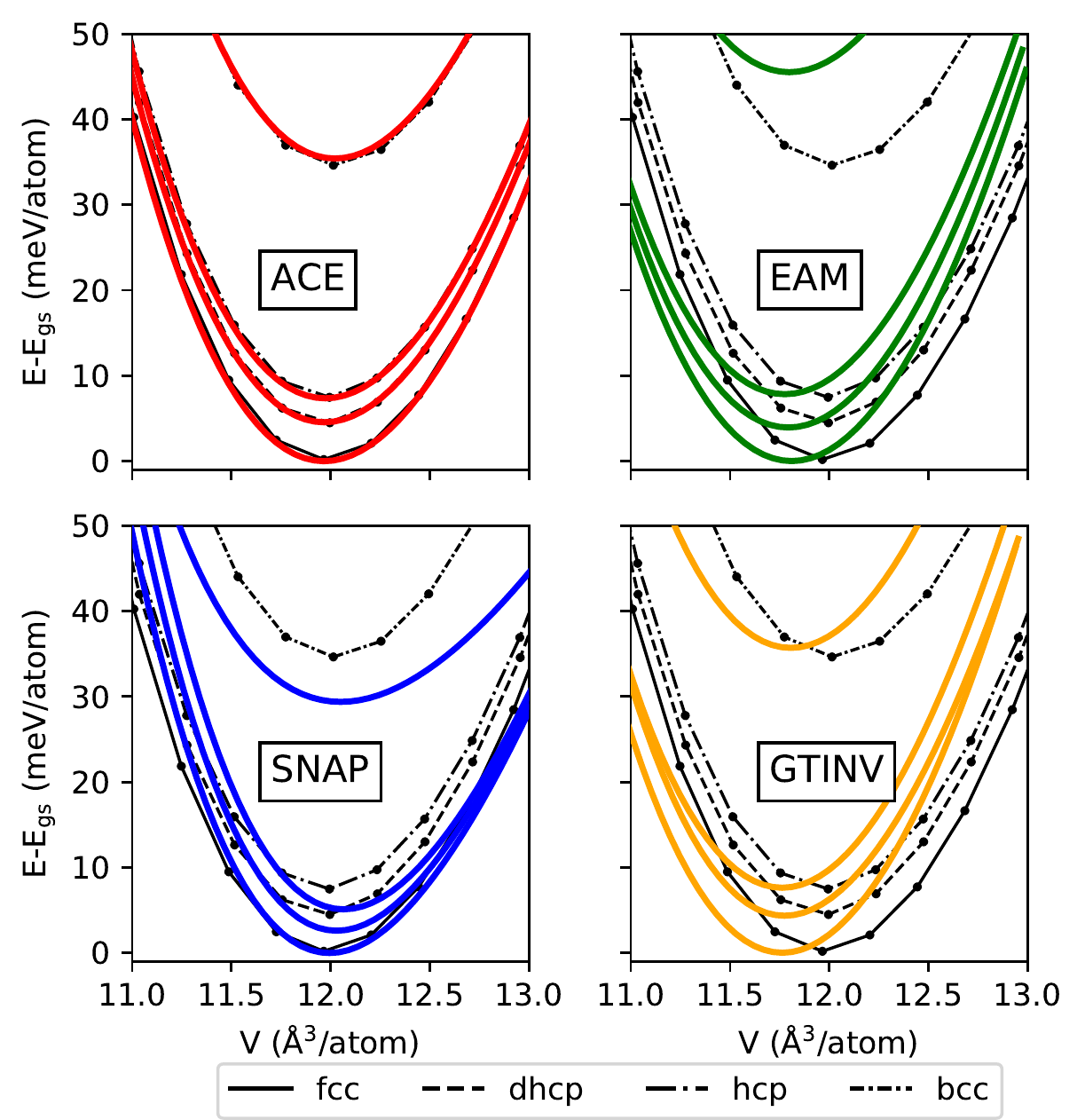}
    \caption{{\bf Energy vs. volume for low-energy bulk phases of Cu.} Lines with symbols correspond to DFT reference data.}
    \label{fig:Cu_E_V_zoom}
\end{figure}

\paragraph{Structural energy differences}
A detailed analysis of structures that are energetically close to the fcc ground state is presented in Fig.~\ref{fig:Cu_E_V_zoom}. All potentials reproduce correctly the  structural order of fcc $\to$ dhcp $\to$ hcp $\to$ bcc. The energy minima predicted by EAM and GTINV potentials are shifted to smaller volumes, which may be due to different DFT reference data. EAM and SNAP also show larger discrepancies for the fcc-bcc energy difference. 

\paragraph{Elastic moduli}
The elastic moduli for the ground state fcc structure are summarized in Tab.~\ref{tab:Cuoverview}. ACE and EAM reproduce the DFT reference very well, while small deviations are observed for SNAP and slightly larger for GTINV. Similar outcomes are obtained for other bulk phases (see SI) with ACE and EAM giving consistently the best agreement with DFT.

\begin{table}[]
\center{
\caption{{\bf Basic properties of Cu.} DFT reference is FHI-aims unless noted otherwise. \label{tab:Cuoverview}}
\label{tab:Cu_basic_prop}
\begin{tabular}{l|c|c|c|c|c|c|c}
\hline
\hline
 &  ACE &  EAM & GTINV & SNAP &&  DFT & Exp. \\
\hline
\hline
\multicolumn{8}{l}{Elastic moduli (GPa)}  \\
\hline
$C_{11}$  &  177 &  176 &  188 &  176 &&  177 &  177\cite{Ledbetter1981elastic} \\
$C_{12}$  &  132 &  133 &  141 &  143 &&  132 & 125\cite{Ledbetter1981elastic} \\
$C_{44}$  &   82 &   82 &   88 &   88 &&   82 & 81\cite{Ledbetter1981elastic} \\
\hline
\hline
\multicolumn{8}{l}{Surface energies $\gamma_\textrm{surf}$ (J/m$^2$)} \\
\hline
$(111)$              &  1.36 &  1.24 &  7.81 &  1.29 &&  1.36 & - \\
$(100)$              &  1.51 &  1.35 &  9.50 &  1.48 &&  1.51 & - \\
$(110)$              &  1.59 &  1.48 &  7.82 &  1.56 &&  1.57 & - \\
\hline
\hline
\multicolumn{8}{l}{Vacancy formation/migration, interstitial formation (eV)} \\
\hline
E$_\mathrm{vac}^\mathrm{f}$ &  1.12 &  1.28 &  1.13 &  1.37 &&  1.07 & 1.27\cite{Siegel1978vacancy} \\
E$_\mathrm{vac}^\mathrm{m}$ &  0.71 &  0.69 &  0.78 & 0.91 &&  0.74 & 0.67-0.76\cite{ehrhart1991atomic} \\
I$_\mathrm{db(100)}$ &  3.14 &  3.12 &  3.25 &  2.80 && 3.10\cite{Ma2021} & 2.8-4.2\cite{Ullmaier1991properties}\\
I$_\mathrm{oct}$     &  3.38 &  3.29 &  3.48 &  2.96 && 3.35\cite{Ma2021}  & - \\
I$_\mathrm{tetr}$    &  3.68 &  3.63 &  3.87 &  3.42 && 3.64\cite{Ma2021}  & -\\
\hline
\hline
\multicolumn{8}{l}{Stacking fault energies (mJ/m$^2$)} \\
\hline
$\gamma_\mathrm{ESF}$    &   48 &   45 &   -679 &    29 &&   43 & - \\
$\gamma_\mathrm{ISF}$    &   43 &   45 &  -1101 &    29 &&   41 & - \\
$\gamma_\mathrm{MAX}$    &  826 &  771 &   1635 &  1139 &&  826 & - \\
$\gamma_\mathrm{MIDDLE}$ &  500 &  479 &    510 &   562 &&  494 & - \\
$\gamma_\mathrm{TWIN}$   &   22 &   22 &   -628 &    14 &&   21 & - \\
\hline
\hline
\multicolumn{8}{l}{Melting temperature  (K)} \\
\hline
T$_\mathrm{M}$ &  1272    &  1324   &  -    &  1372  &&   1251\cite{Zhu2017} & 1358\cite{chekhovskoi2000calorific}    \\
\hline
\end{tabular}
}
\end{table}

\begin{figure}
    \includegraphics[width=\columnwidth]{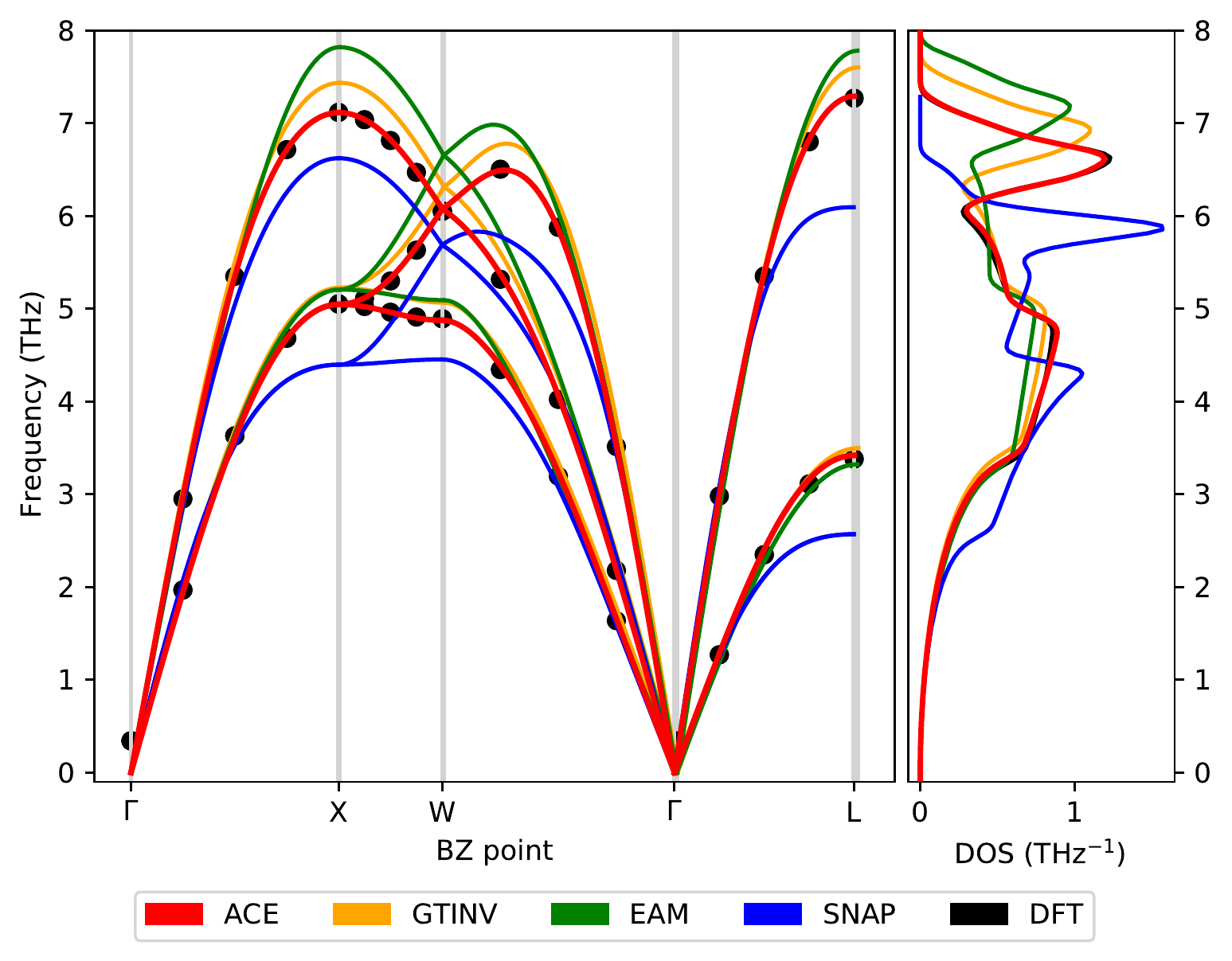}
    \caption{{\bf Phonon properties for fcc Cu.} Phonon band structure (left) and density of states (right).}
    \label{fig:Cu_fcc_phonon}
\end{figure}

\paragraph{Phonons}
Figure~\ref{fig:Cu_fcc_phonon} shows a comparison of phonon band structures and densities of states (DOS) for fcc Cu. Despite not having fitted any phonon frequencies explicitly, ACE provides the best match to the reference DFT data. The EAM and GTINV potentials overestimate the width of the DOS , while SNAP underestimates it. These conclusions apply also for the phonon DOS of other crystal structures that are shown in the {SI}.

\begin{figure}
    \includegraphics[width=\columnwidth]{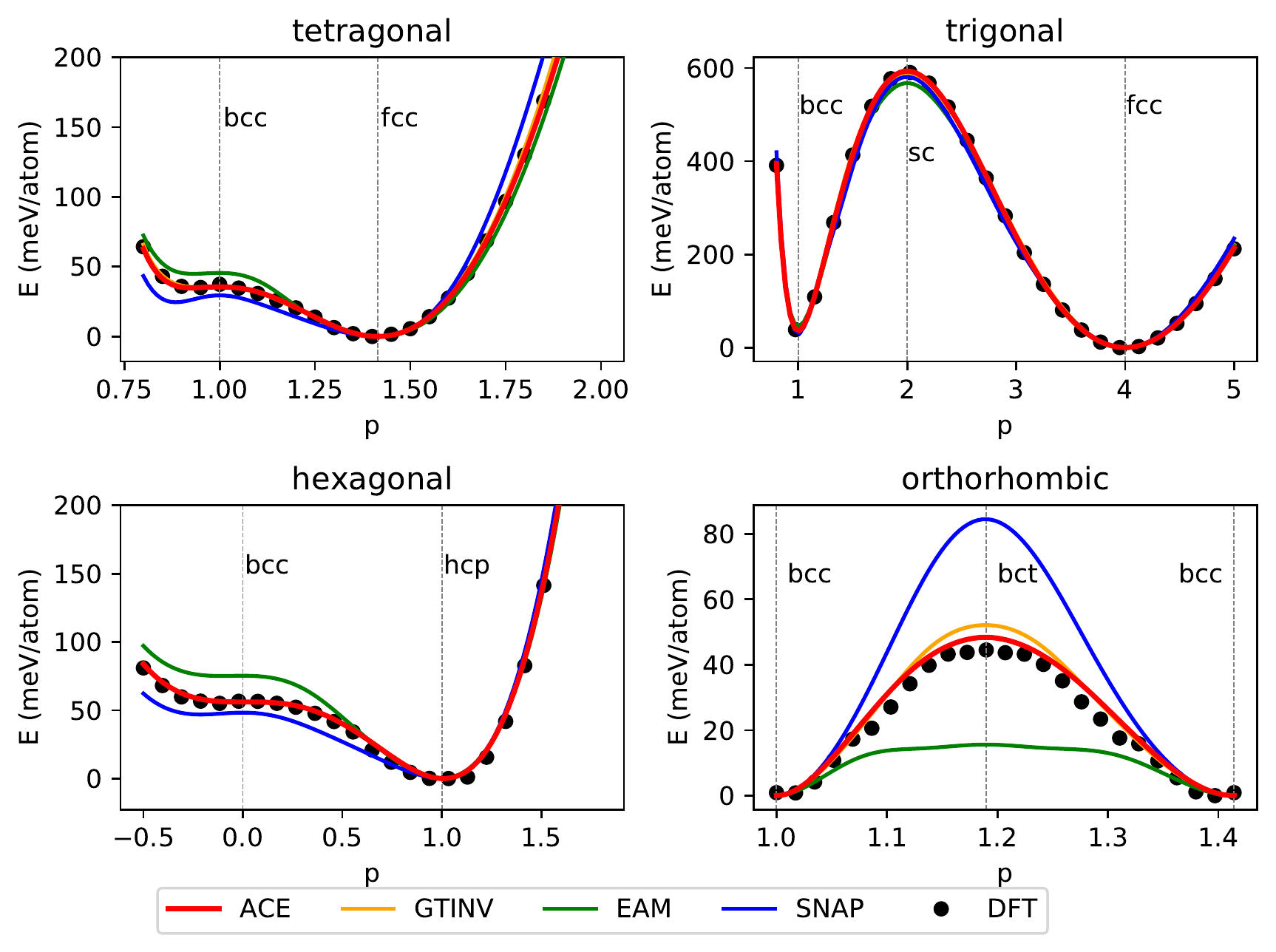}
    \caption{{\bf Transformation paths for Cu.}}
    \label{fig:Cu_fcc_trans_path}
\end{figure}

\paragraph{Structural transformations}
Transformations between different crystal structures present a sensitive test for any interatomic potential as both bond distances and bond angles are changed simultaneously.  In addition, the associated changes in atomic coordination effectively scrutinize the screening of pairwise terms by many-atom contributions. As shown in Fig.~\ref{fig:Cu_fcc_trans_path}, all potentials agree well with the reference DFT data for the tetragonal, trigonal and hexagonal paths. However, only ACE provides an excellent quantitative interpolation for all structures along all considered transformation paths.  Especially, the orthorhombic transformation, which can be regarded as a generalization of the Bain path~\cite{luo2002ideal}, is challenging for the other potentials.

\paragraph{Melting transition and thermal expansion}

We used thermodynamic integration to evaluate the free energy of the solid and liquid phases of Cu. The free energies intersect at $T=1272$~K, about 20~K above the $1251 \pm 15$~K predicted by DFT \cite{Zhu2017}. The EAM and SNAP melting temperature at $T=1325$~K and {$1372$~K}  are close to the experimental value of $1358$ K. The prediction of the melting point with GTINV was not possible due to long evaluation times and the lack of a parallel implementation.

\begin{figure}
    \includegraphics[width=0.75\columnwidth]{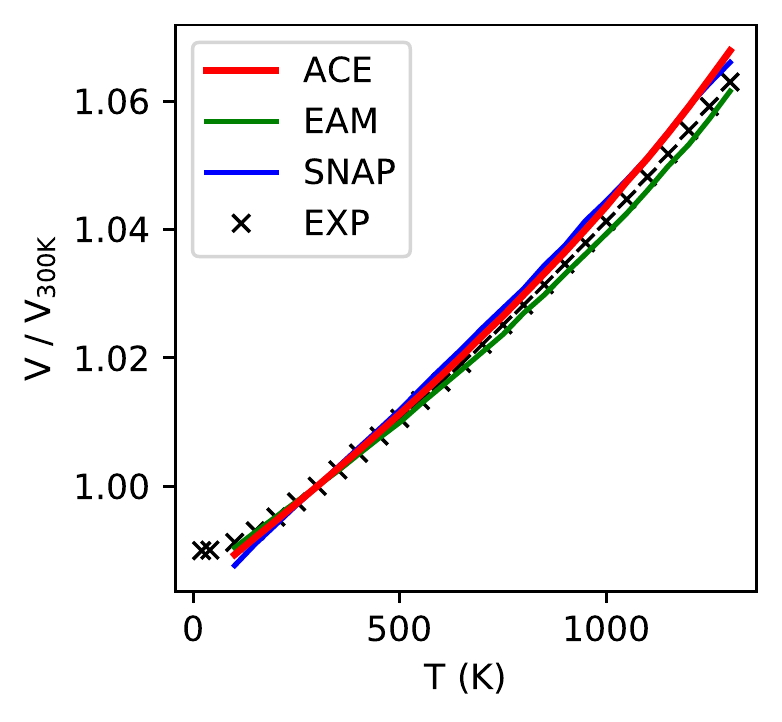}
    \caption{{\bf Thermal expansion for fcc Cu.} Experimental data are taken from Ref.~\onlinecite{Wang1996thermal}.}
    \label{fig:Cu_thermal_exp}
\end{figure}

Fig.~\ref{fig:Cu_thermal_exp} shows the thermal expansion as obtained from MD simulations in the NPT ensemble. All models agree well with the experimental data for temperatures up to 600 K and exhibit minor deviations at high temperatures. 

\subsubsection{Interfaces and surfaces}

Planar defects include internal interfaces, such as stacking faults (SF) and grain boundaries (GBs), where the local atomic density does not vary significantly but bond angles change compared to bulk.  In contrast, at free surfaces the bond angles remain mostly unaltered but the surface atoms loose about half of their neighbors. Typically, central-force models 
such as EAM provide a good description of structures and energies of GBs but cannot capture well the large local density changes at surfaces which usually leads to underestimation of surface energies. 

\paragraph{Stacking faults}
The small energy differences between the close-packed fcc, hcp and related structures in Cu imply small SF energies. ACE predicts the SF energies in very good agreement with DFT reference data, as shown in Tab.~\ref{tab:Cuoverview}, with comparable predictions from EAM. GTINV predicts negative stacking fault energies, hinting at a different ground state. SNAP provides stacking fault energies with slightly larger deviations from the reference data.

\begin{figure}
    \includegraphics[width=0.65\columnwidth]{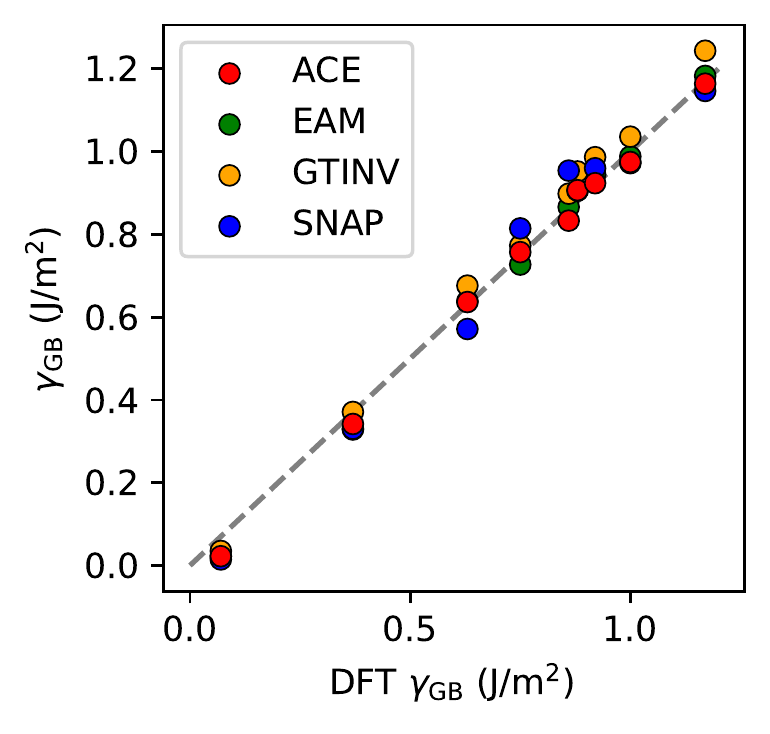}
    \caption{{\bf Cu twin and twist GBs.} Energies with respect to DFT reference from~\citet{Zheng2020grain} {See {SI} for more details.} }
    \label{fig:Cu_fcc_GB}
\end{figure}

\paragraph{Grain boundaries}
The energies of several twin and twist symmetric GBs ($\Sigma = 3,5,9$) are compared in Fig.~\ref{fig:Cu_fcc_GB} to reference DFT data from the Materials Project database~\cite{Zheng2020grain}  (see SI for more details). As expected, all potentials predict the GB energies very accurately, which suggests good transferability of all models for environments with small local density variations.

\begin{figure}
    \includegraphics[width=0.65\columnwidth]{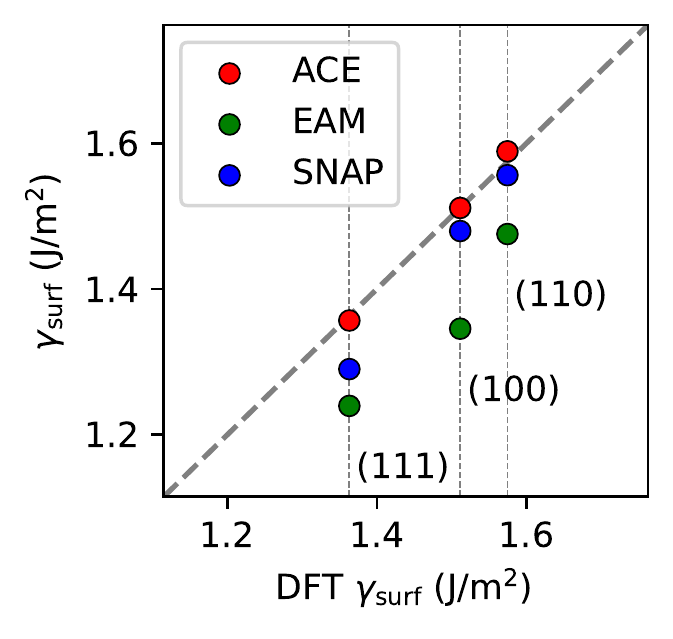}
    \caption{{\bf Cu surface energies.} 
    }
    \label{fig:Cu_fcc_surface_energies} 
\end{figure}

\paragraph{Surface energies}
As noted above, surfaces present a much more stringent test than GBs. ACE provides the best agreement with DFT reference data for all low-index surfaces, as shown in Fig.~\ref{fig:Cu_fcc_surface_energies},  while both SNAP and EAM consistently underestimate the surface energies. For GTINV we observed a detachment of the top surface layers during relaxation which resulted in unphysically high surface energies that were excluded from the comparison.

\begin{figure}
    \includegraphics[width=0.99\columnwidth]{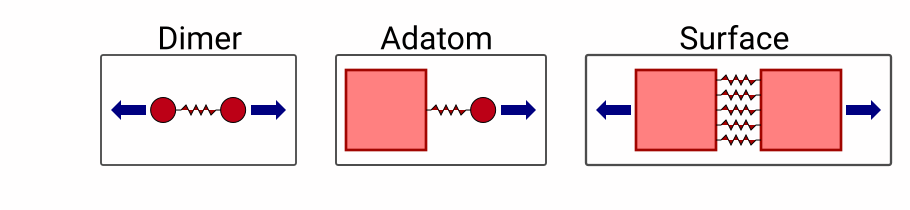}\\
    
    \includegraphics[width=\columnwidth]{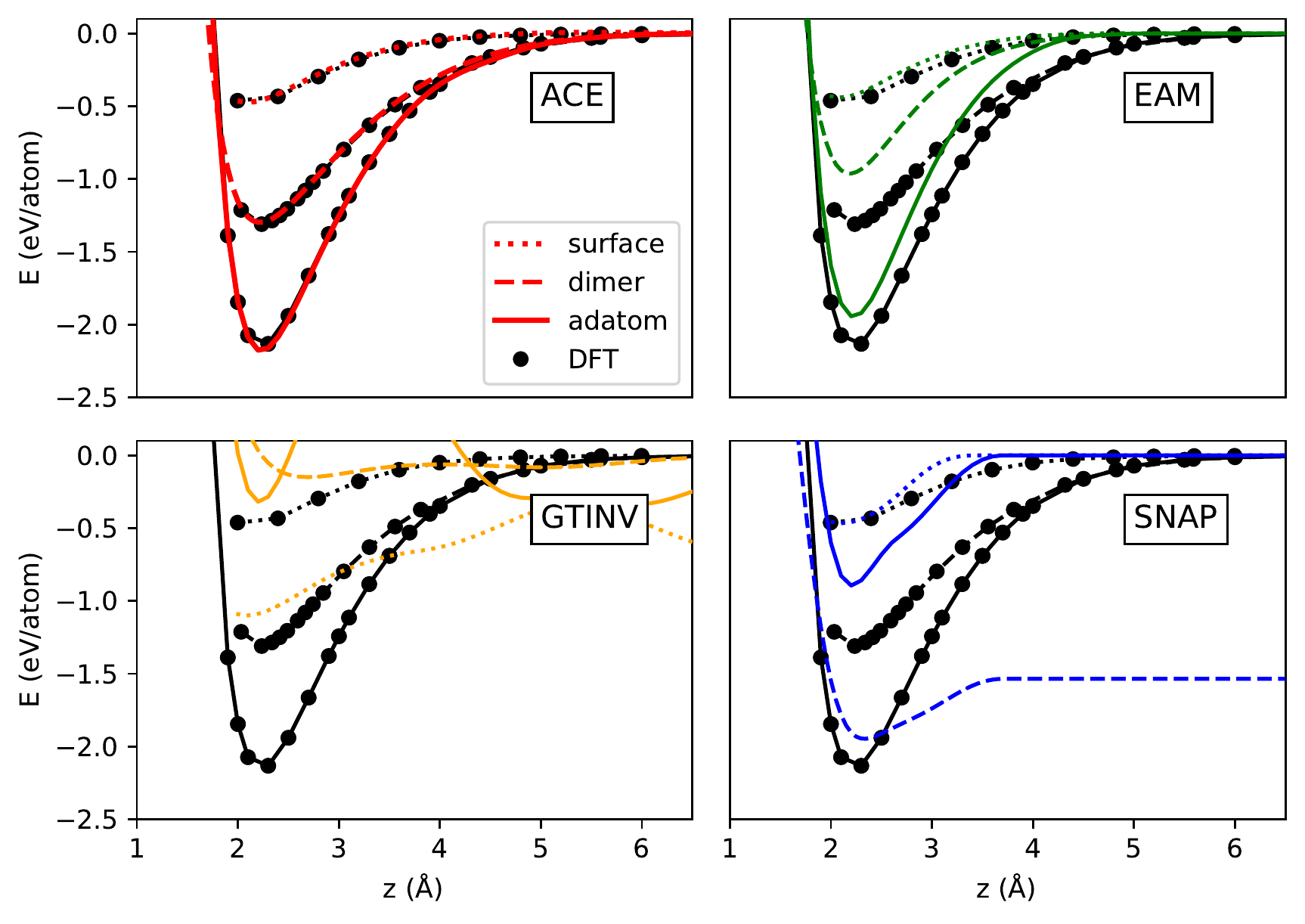}
    \caption{{\bf Decohesion in different environments.} Comparison of Cu dimer dissociation to a detachment of the "on-top" Cu adatom from the $(111)$ surface and a rigid decohesion of two Cu bulk slabs along the $\langle 111 \rangle$ direction.}
    \label{fig:Cu_fcc_surface_dimer_adatom} 
\end{figure}

\paragraph{Bond breaking}
In addition to the energetics of surfaces, we examined bond breaking in various atomic environments. Such tests have practical relevance as they are related to fracture, surface adsorption or vaporization. We designed three distinct decohesion tests that are schematically shown in Fig.~\ref{fig:Cu_fcc_surface_dimer_adatom}. These tests compare bond dissociation in the Cu dimer, detachment of a Cu adatom from the $(111)$ surface, and an ideal rigid decohesion of bulk Cu slabs that leads to the formation of two $(111)$ free surfaces.  As can be seen from Fig.~\ref{fig:Cu_fcc_surface_dimer_adatom}, ACE is the only model that is able to describe quantitatively accurately the impact of the atomic environment on bond breaking. The presence of neighboring atoms leads to an effective screening of the interatomic bonds and their interaction ranges~\cite{Nguyen-Manh00}. The dimer and the adatom have no neighbors so that their interaction range is longer than the interaction between two surfaces whose atoms are surrounded by bulk. Given the simplicity of EAM, it provides a surprisingly good account of bond breaking in the very different environments, while GTINV and SNAP have problems with this test.

\subsubsection{Point defects, small clusters and 2d structures}

\paragraph{Point defects}
Properties of point defects, such as mono-vacancy and self-interstitial, are often included in the fitting dataset.  Given that only unrelaxed vacancy configurations were part of the reference data, ACE reproduces the vacancy formation energy very well while the other potentials overestimate the DFT reference by 0.1-0.3 eV, see Tab.~\ref{tab:Cuoverview}. The migration barrier is reproduced well, too, by the models, apart from SNAP that overestimates the barrier.  

None of the interstitial configurations were included in the ACE training set, but ACE results are consistent with those of the other potentials and together with EAM agree best with recent DFT results.\cite{Ma2021} The $\langle 100 \rangle$ dumbell is predicted to have the lowest energy, followed by the octahedral and tetrahedral configurations. These predictions are consistent with those for other fcc metals.\cite{Connetable2015} 

\paragraph{Small clusters}
Small metallic clusters, important for catalysis and nanotechnology, usually form a large number of isomers with energies and structures often governed by subtle electronic structure contributions. For this reason, the predictions of the detailed energetics and structural stability is very challenging for interatomic potentials that are typically aimed at the description of bulk systems.  We compared the predictions of ACE and the other models for three- and four-atomic clusters. 

For the Cu trimer, the ground state structure is an isosceles triangle configuration while the linear trimer corresponds to an energy saddle point and is not dynamically stable~\cite{Cogollo2015}.
In fact, the linear trimer transforms to a metastable configuration of a bent molecule with an obtuse angle of ~130$^\circ$. ACE is the only model that correctly reproduces the instability of the linear trimer and the existence of the metastable bent configuration. EAM predicts the equilateral triangle as the only stable configuration while for SNAP and GTINV both the linear trimer and the equilateral triangle are stable configurations.  The energy differences between the configurations are also reproduced most accurately by ACE while the other models either significantly underestimate (GTINV) or overestimate (EAM, SNAP) the DFT values.

In the case of the tetramer, only ACE and GTINV give correctly the planar equilateral rhombus  \cite{Cogollo2015} as the ground state, albeit GTINV shows also additional metastable configurations.
Both EAM and SNAP favor incorrectly the close-packed tetrahedron which may originate from the lack of or weak angular contributions.

\paragraph{2D structures}

\begin{figure}
    \includegraphics[width=\columnwidth]{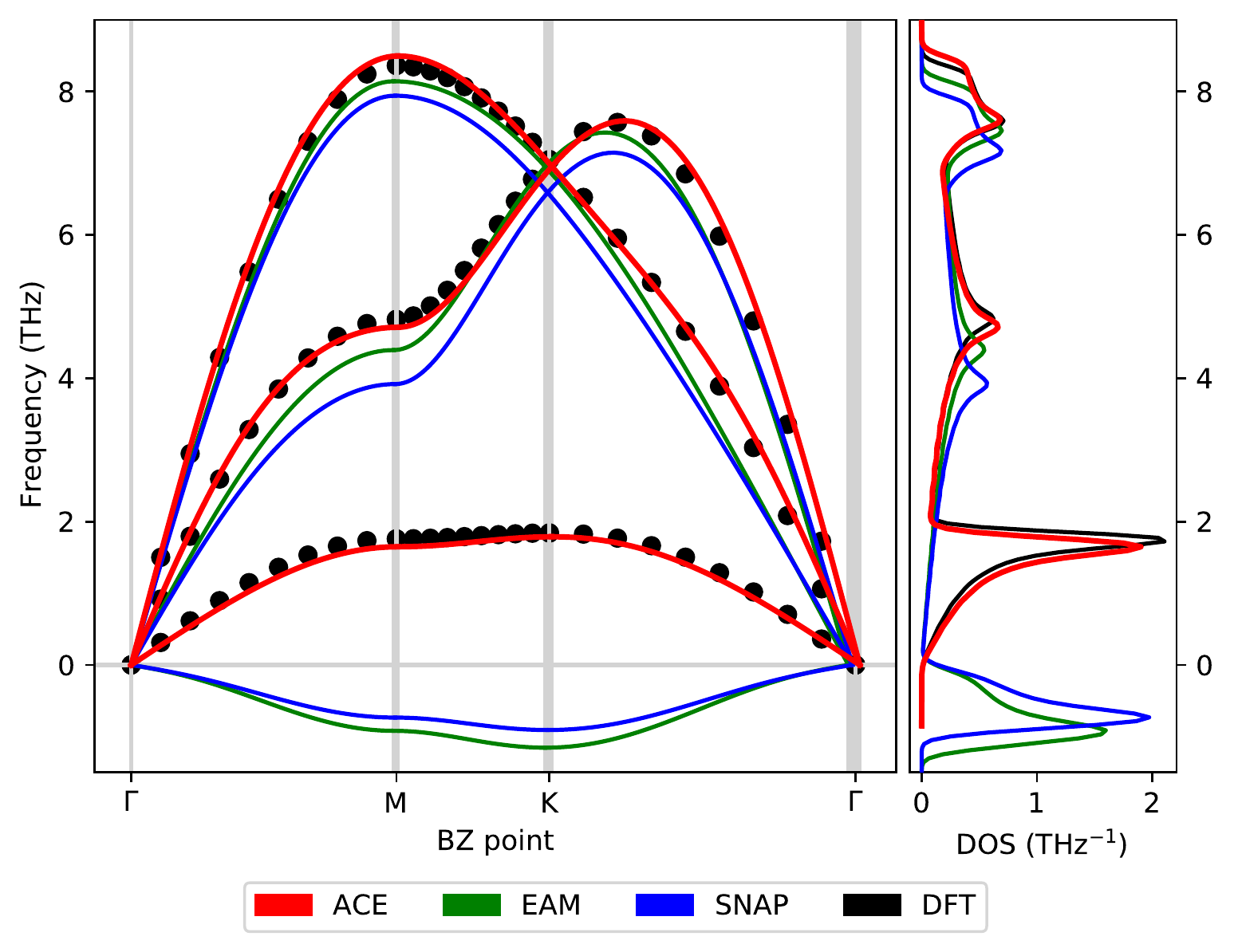}
    \caption{{\bf Phonons in 2D hcp lattice.} Comparison of phonon band structure and phonon DOS for the free-standing hcp Cu monolayer.
    }
    \label{fig:Cu_2D_hex_phonon_dispersion_DOS_supercell5} 
\end{figure}

Planar 2D structures belong to a family of structures that is usually not included in the validation of interatomic potentials for bulk metals. It has been found recently that Cu is the only metal whose free-standing monolayers arranged in honeycomb, square and hexagonal close-packed lattices are dynamically stable~\cite{ono2020dynamical}. 
We investigated in detail the 2D hcp lattice; both EAM and SNAP potentials show dynamic instabilities related to out-of-plane atomic displacements for the $3  \times 3 \times 1$ supercell that we used in our calculations (Fig.~\ref{fig:Cu_2D_hex_phonon_dispersion_DOS_supercell5}).
In contrast, DFT and ACE predict real phonon frequencies that confirm excellent transferability of ACE once more. We note that the 2D hcp structure could not be stabilized using the GTINV potential.

\subsection{Silicon}

The ACE for silicon was created by fitting to an extensive database first introduced to create a general-purpose GAP model for silicon \cite{Bartok18}. This GAP was shown to describe silicon accurately and to also be a {\em qualitatively} better interatomic potential than all other models tested, each best in their class: Stillinger-Weber \cite{Stillinger85,Stillinger86}, EDIP \cite{Justo98}, Tersoff \cite{Tersoff88}, MEAM \cite{Baskes92}, DFTB \cite{Porezag95} and ReaxFF \cite{Buehler06}. In this paper we show that the silicon ACE potential achieves the same accuracy as the GAP model, while being around 30 times faster in evaluation time and also better at extrapolating to unseen configurations.

The following section presents a benchmarking of the ACE silicon potential on a wide range of properties including bulk, surface, liquid and amorphous properties as well as a random structure search (RSS) \cite{Pickard11} test.

\subsubsection{Bulk properties}

\begin{figure*}
    \centering
  \includegraphics[width=0.49\textwidth]{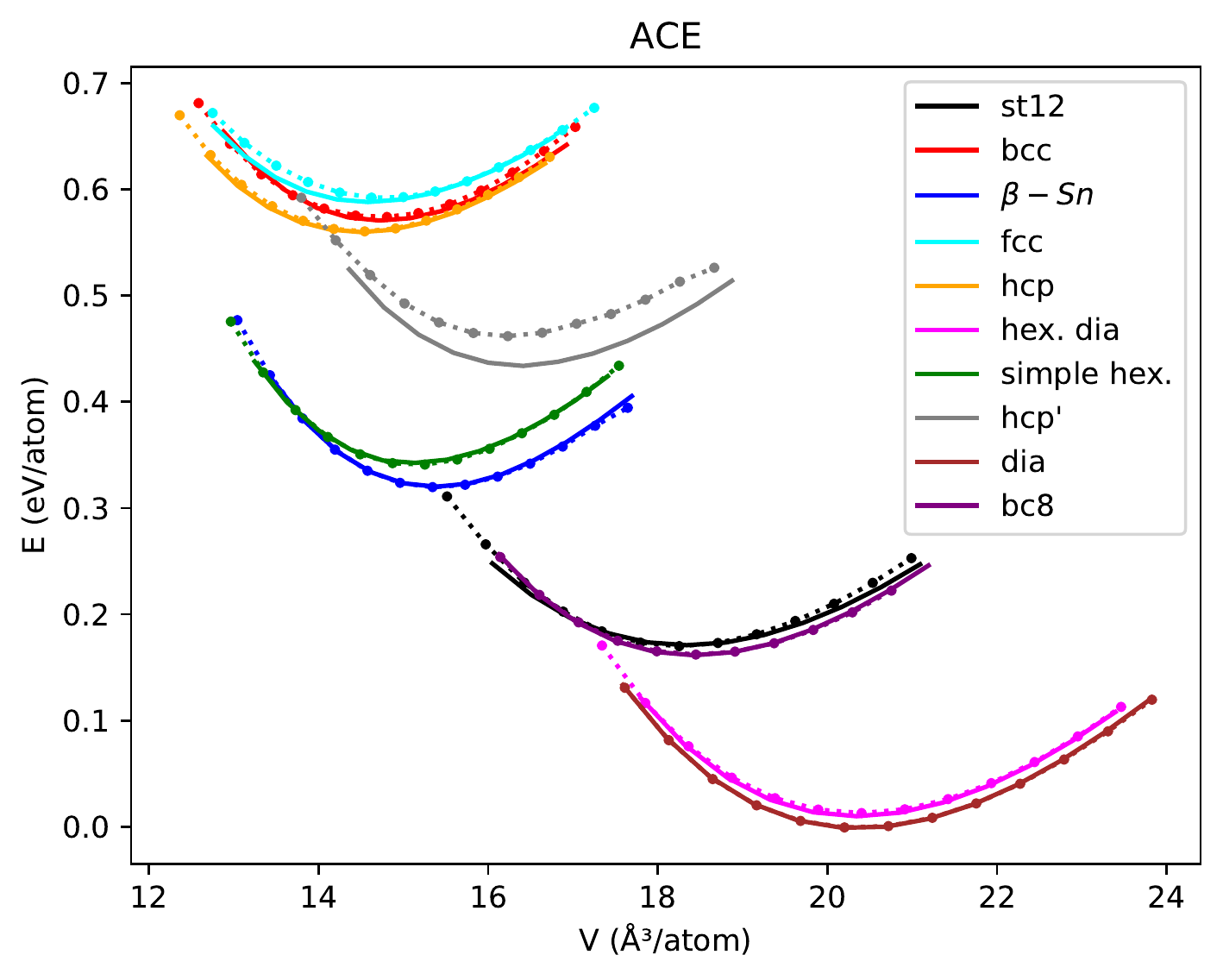}
  \includegraphics[width=0.49\textwidth]{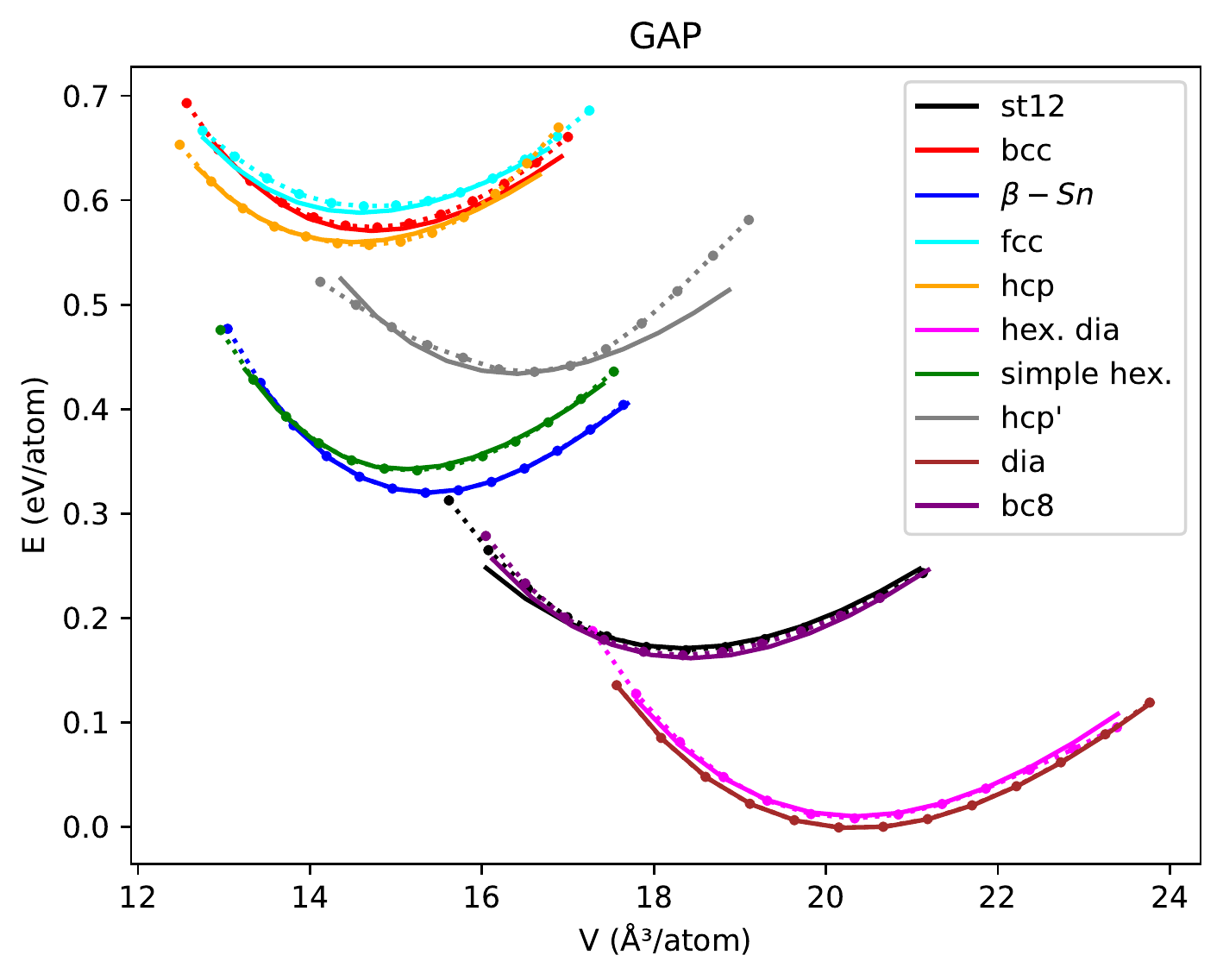}
  \caption{{\bf Energy vs volume for Si.} Shown are bulk crystal lattices compared to DFT reference (solid). Both models not explicitly fitted to the hcp' structure, requiring the ACE (left) and GAP (right) to extrapolate.
  }
    \label{fig:si_bulks_fig}
\end{figure*}

\paragraph{Structural energy differences}
The energies of the diamond, hexagonal diamond, $\beta$-Sn, bc8, st12, bcc, fcc, simple hexagonal (sh), hcp and hcp' are compared to DFT in Fig.~\ref{fig:si_bulks_fig}. Excellent agreement with the DFT reference is observed for all structures apart from hcp'. Si hcp has two minima\cite{Bartok18}, the conventional hcp with $c/a \approx \sqrt{3/2}$, and hcp' with $c/a < 1$. The hcp' crystal structure is not contained in the DFT reference silicon database, however, both ACE and GAP predict the minimum. The GAP predicts the DFT reference energy at the minimum more accurately than ACE, while the latter gives a better estimate of the curvature.

\begin{figure}
    \includegraphics[width=\columnwidth]{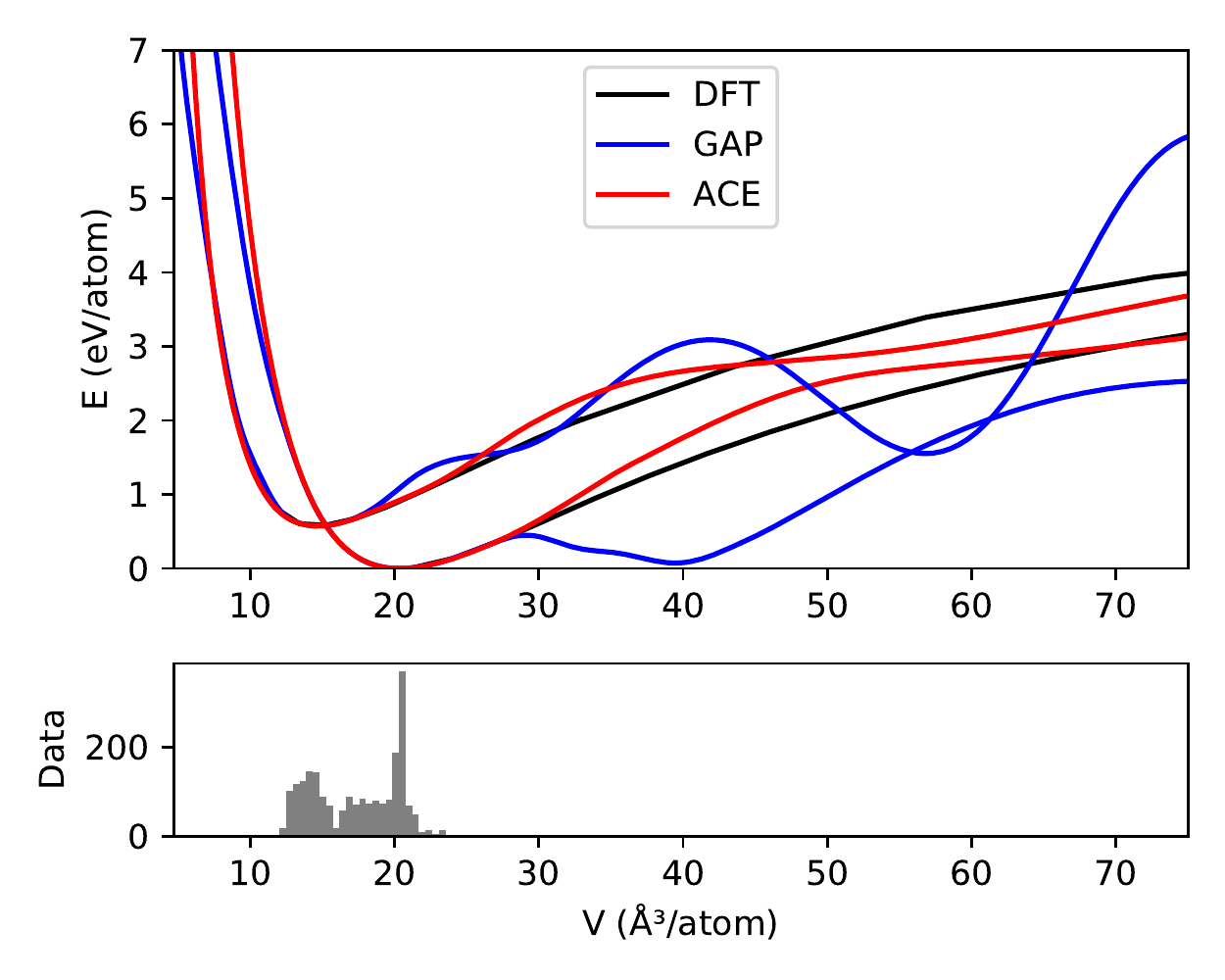}
    \caption{{\bf Extrapolation for large volumes.} Comparison of ACE and GAP to DFT for the energy-volume curves for diamond and bcc structures of Si. The lower panel shows the volume distribution of the reference data.
    }
    \label{fig:si_extrap_bulks_fig}
\end{figure}

The energy versus volume curves for the silicon diamond and bcc are extended over a wide volume range in Fig.~\ref{fig:si_extrap_bulks_fig}. Both potentials accurately describe the minima around 15 and 20 \AA$^{3}$/atom for bcc and diamond, as previously shown in Fig.~\ref{fig:si_bulks_fig}. At larger volumes GAP exhibits unphysical high-energy minima. ACE does not show these minima and is close to the DFT reference, demonstrating better extrapolation compared to GAP. This extrapolative behavior is remarkable since there is no reference data at these large volumes as shown by the data density in the lower panel. 

\paragraph{Elastic moduli}
The elastic constants for Si in the diamond structure are summarized in Tab.~\ref{tbl:si_table}. Both ACE and GAP match the DFT reference within a few percent.

\begin{table}[]
\centering
\begin{tabular}{l|c|c|c}
\hline
\hline
 &  ACE &  GAP  &  DFT \\
\hline
\hline
\multicolumn{4}{l}{Elastic moduli (GPa)} \\
\hline
$B$  &  80 &  83 &  82 \\
$C_{11}$  &  142 & 145 &   147 \\
$C_{12}$  &  50 & 52 &  50 \\
$C_{44}$  &   70 &  69 &    73 \\
\hline
\hline
\multicolumn{4}{l}{Surface energies $\gamma_\textrm{surf}$ (J/m$^2$)} \\
\hline
(111)              &  1.47 &  1.50 &   1.56 \\
(100)              &  2.11 &  2.12 &   2.17 \\
(110)              &  1.51 &  1.55 &   1.52 \\
\hline
\hline
\multicolumn{4}{l}{Vacancy/interstitial (eV)} \\
\hline
E$_\mathrm{vac}$ &   3.72 &  3.73 &    3.67 \\
I$_\mathrm{db}$ &  3.55 &  3.59 &  3.66 \\
I$_\mathrm{hex}$     &  3.36 &   3.48 &  3.72 \\
I$_\mathrm{tetr}$    &  3.47 &    3.58 &  3.91 \\
\hline
\end{tabular}\\
\caption{ {\bf Basic properties of Si}. Shown are elastic moduli, surface energies, vacancy and interstitial energies in diamond silicon. DFT reference is CASTEP \cite{CASTEP}.}
\label{tbl:si_table}
\end{table}

\begin{figure}
    \includegraphics[width=\columnwidth]{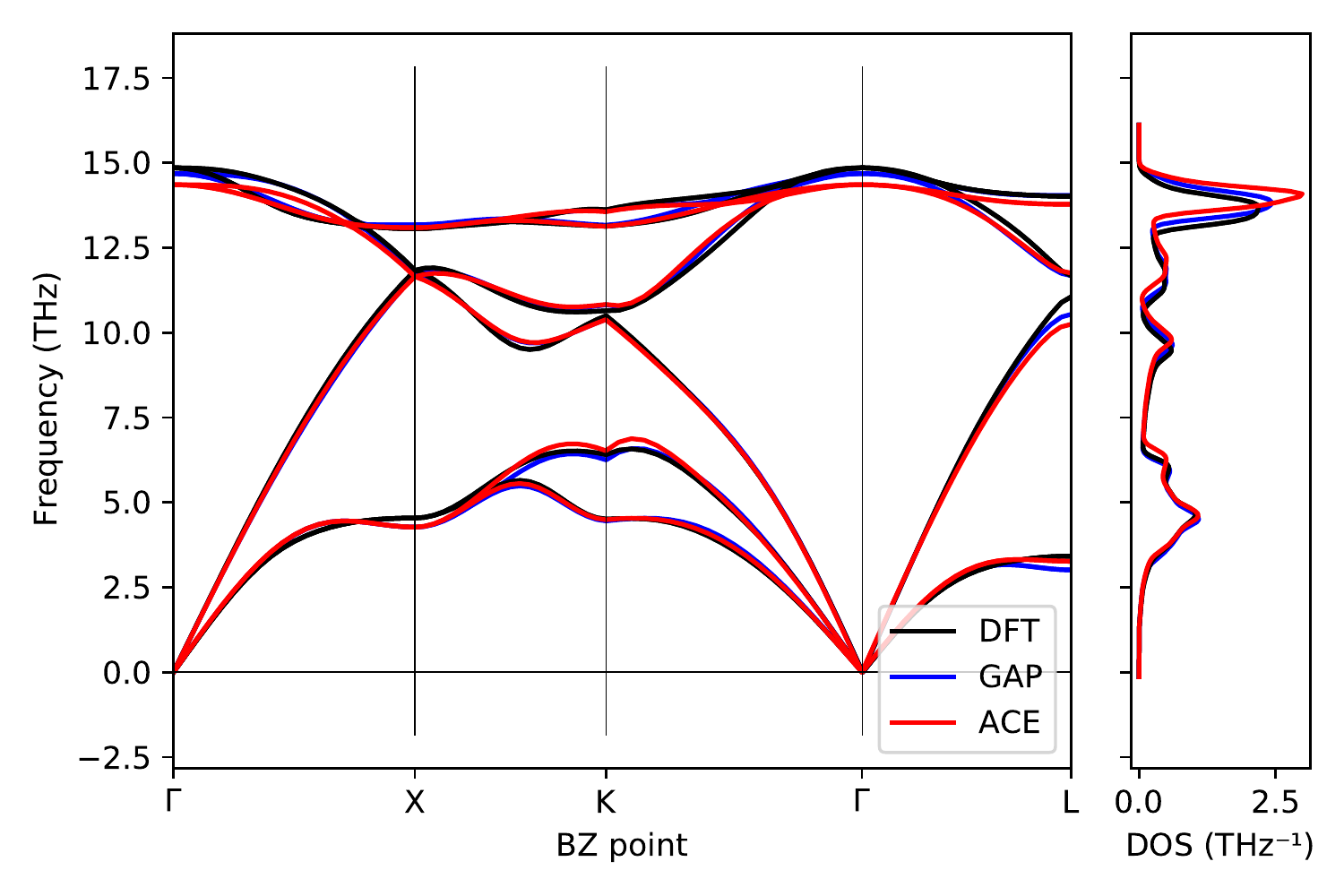}
    \caption{{\bf Phonon dispersion for diamond Si.}}
    \label{fig:si_phonon_fig}
\end{figure}

\paragraph{Phonons}
The phonon dispersion for ACE and GAP is compared against the DFT reference in Fig.~\ref{fig:si_phonon_fig}. Both GAP and ACE accurately describe the phonon spectrum in comparison to the DFT reference, with the band width of GAP showing a better match to DFT. More silicon phonon spectra are found in the SI.

\begin{figure}
    \includegraphics[width=\columnwidth]{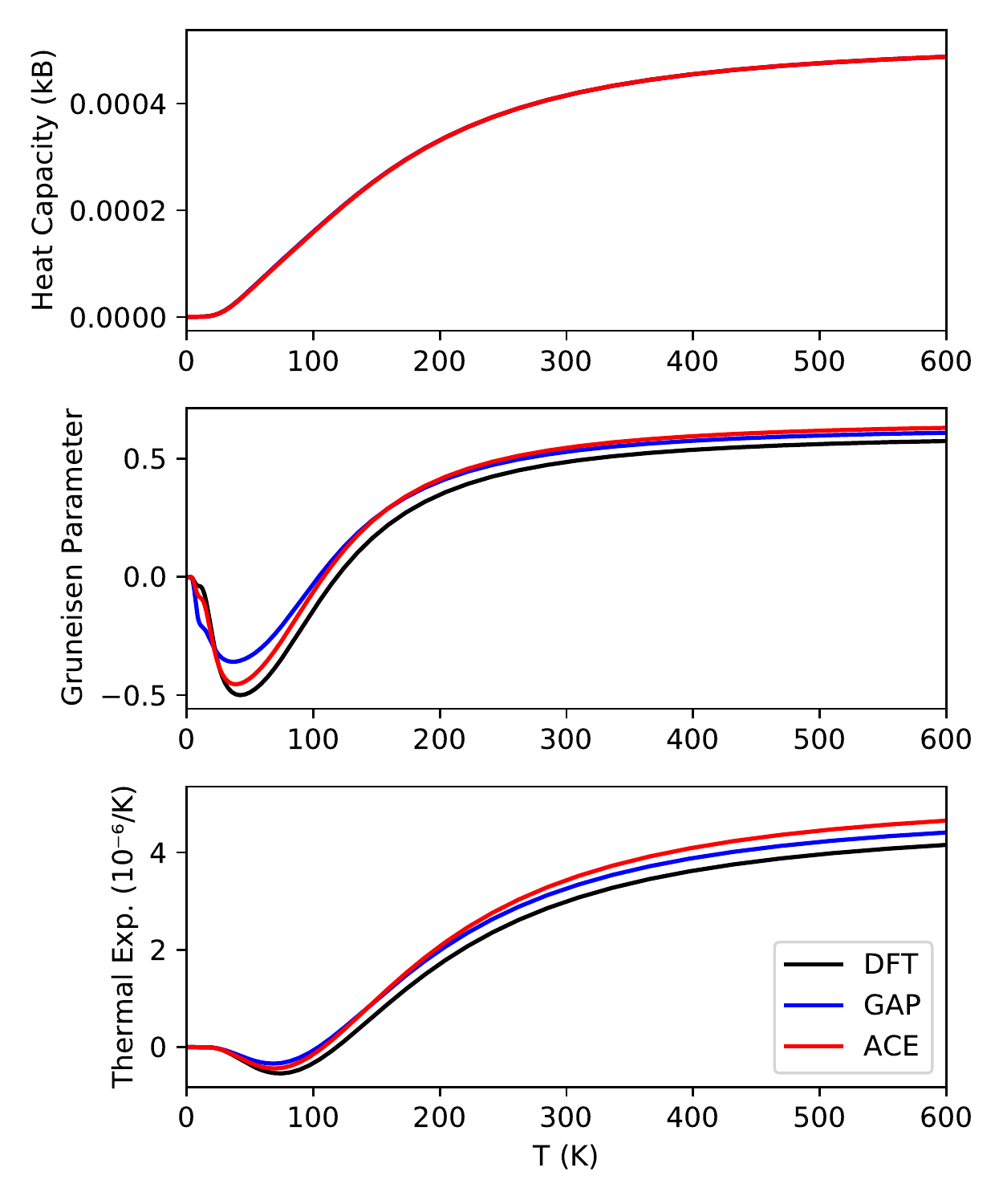}
    \caption{{\bf Thermal properties of diamond Si.} Thermal expansion, Gr\"uneisen parameter and specific heat.}
    \label{fig:si_qha}
\end{figure}

\paragraph{Thermal expansion}
We investigated the thermal expansion, Gr\"uneisen parameter and heat capacity of ACE and GAP in the quasi-harmonic approximation as shown in Fig~\ref{fig:si_qha}. Diamond silicon displays negative thermal expansion at low temperatures \cite{Okada84}, which ACE models very well compared to the DFT reference, and more accurately than GAP. The heat capacity is described with almost perfect agreement, whereas the thermal expansion saturates at a slightly too high value for high temperature (for both GAP and ACE). 

\paragraph{Liquid phase}

\begin{figure}
    \includegraphics[width=\columnwidth]{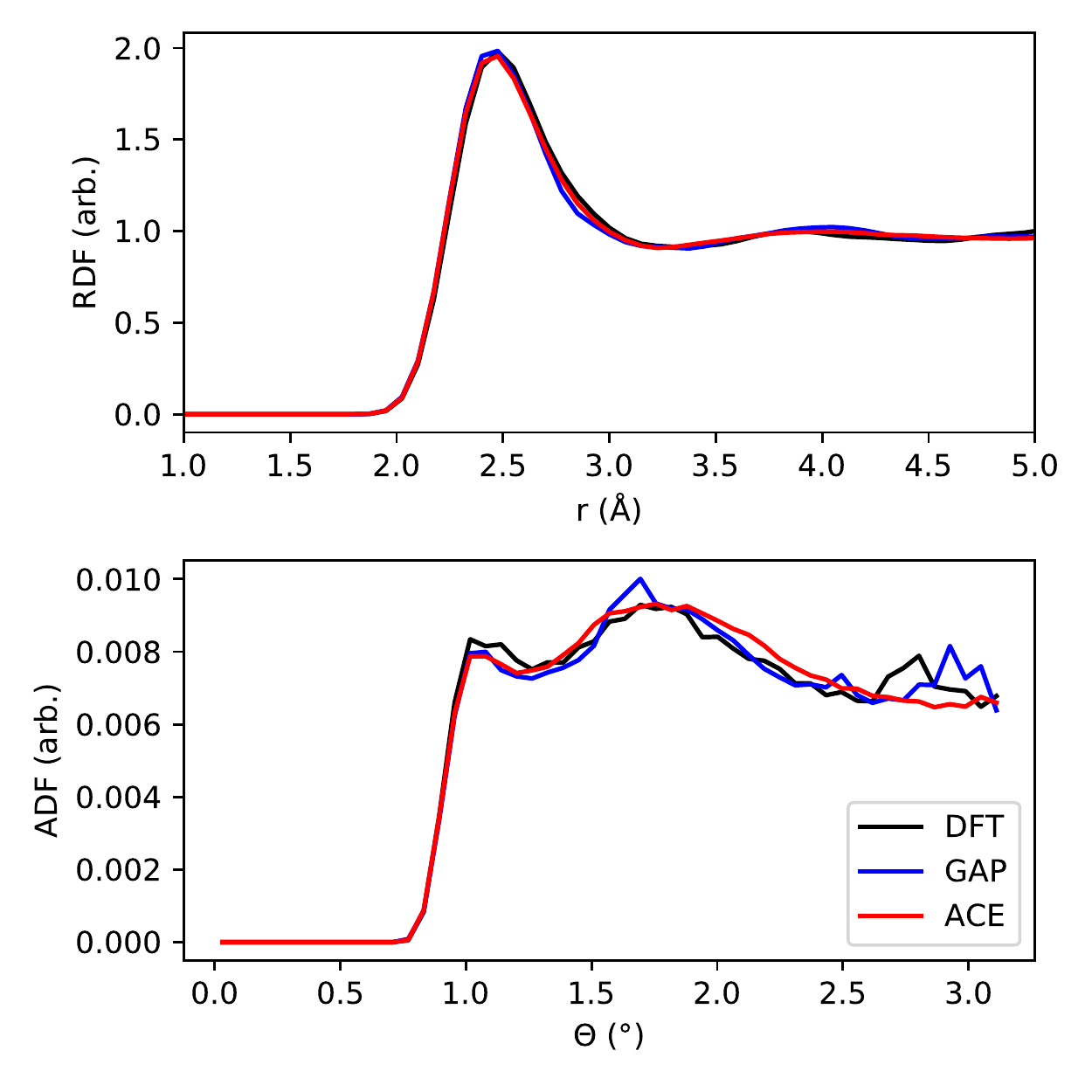}
    \caption{{\bf Structural properties of liquid Si.} Radial distribution function (RDF) (top) and angular distribution function (ADF) (bottom) at P = 0 GPa and T = 2000K.}
    \label{fig:si_liquid}
\end{figure}

ACE was also tested in a liquid simulation on an eight-atom 2x2x2 supercell (64 atom) and compared to GAP and the DFT reference. The radial distribution function (RDF) and angular distribution function (ADF) where averaged over 20000 MD steps (0.25 fs timestep). The DFT reference data was generated using \verb+CASTEP+ averaging over 9700 MD steps (0.25 fs timestep) taken at $T$=2000K, using a 200 eV plane-wave energy cutoff and 0.05 \AA$^{-1}$ k-point spacing. The results are shown in Fig.~\ref{fig:si_liquid} and demonstrate excellent agreement between ACE, GAP and DFT reference. 

\paragraph{Amorphous phase}

\begin{figure}
    \includegraphics[width=\columnwidth]{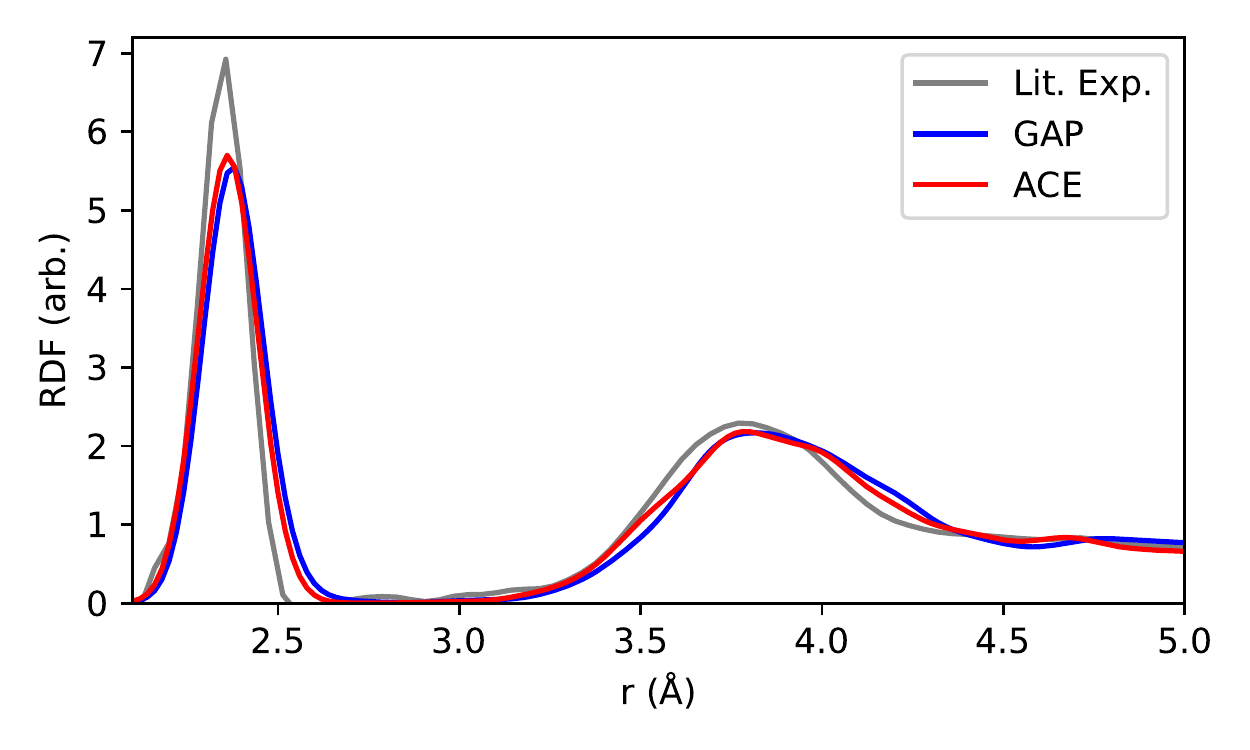}
    \caption{{\bf Structural properties of amorphous Si.} Radial distribution function (RDF) for a 216-atom configuration generated by cooling liquid Si from 2000K to 500K comparing ACE, GAP and experimental data}
    \label{fig:si_amorph}
\end{figure}

Amorphous silicon is a tetrahedrally coordinated phase that forms upon rapid quenching from the melt. Here we quench a 216-atom sample of liquid Si
from 2000 K to 500 K at a rate of $10^{12}$ 
K/s with a 1 fs time step (1.5 $\times$ $10^6$ steps) using the LAMMPS software \cite{LAMMPS}. After the MD steps the final configuration was relaxed to a local minimum with respect to cell size and shape. The radial distribution functions (RDF) of both GAP and ACE are compared to experimental results \cite{Laaziri99} (since DFT results are not computationally feasible) in Fig.\ref{fig:si_amorph}. Both GAP and ACE accurately describe the first and second neighbor peaks, and no atoms in the range ($2.5$ \AA $\leq r \leq 3.25 $ \AA). 

\subsubsection{Surfaces}

\paragraph{Surface energies}
The surface formation energy in the $(100), (110), (111)$ directions are summarized in Table~\ref{tbl:si_table}. ACE and GAP agree very well with the DFT reference.

\begin{figure}
    \includegraphics[width=\columnwidth]{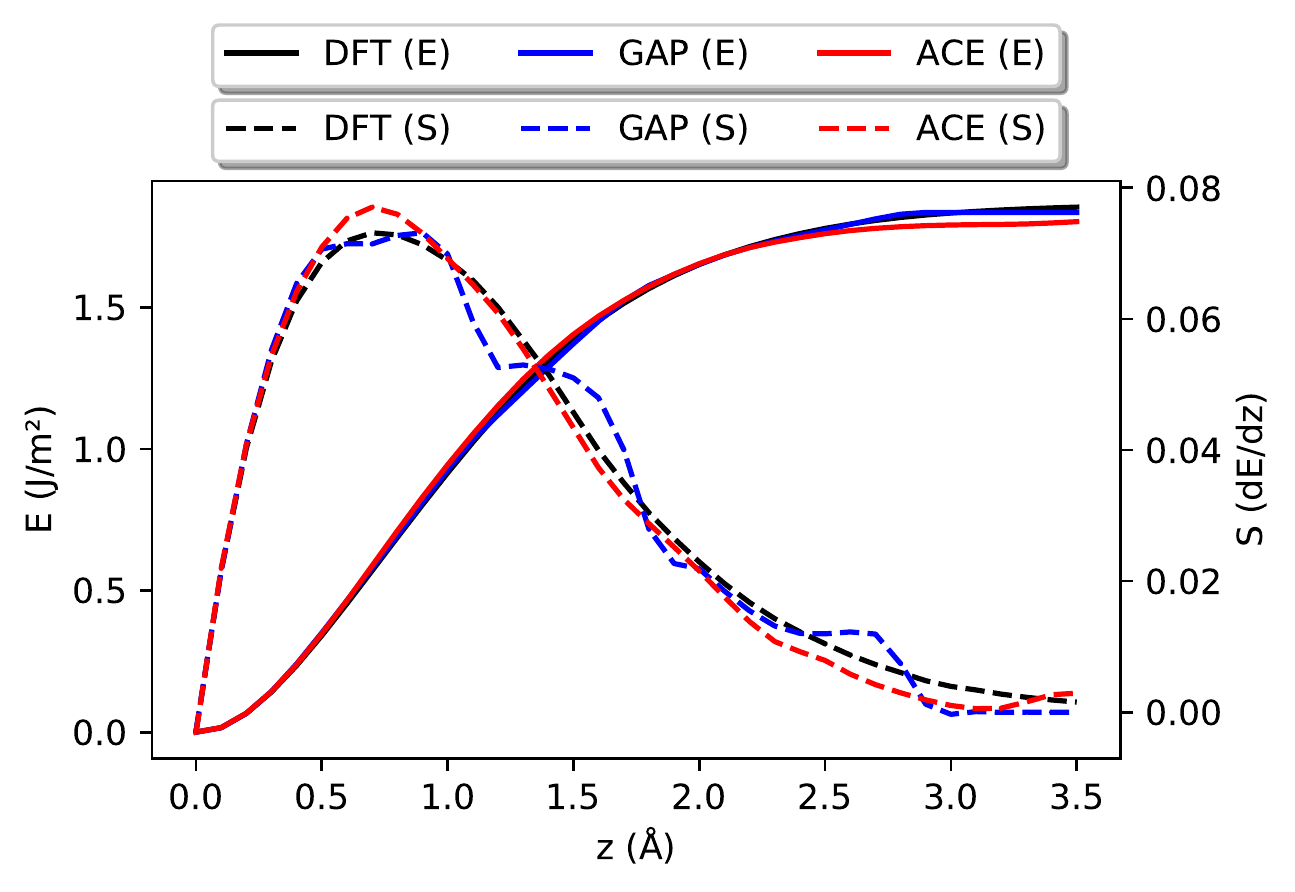}
    \caption{{\bf Rigid decohesion of Si. } Energy (solid) and stress (dashed) curve from bulk crystal to the unrelaxed (110) surface.}
    \label{fig:si_decon_fig}
\end{figure}

\paragraph{Surface decohesion}
Surface decohesion bridges two parts of the training database, from bulk crystal diamond to the unrelaxed ($110$) surface, see Fig.~ \ref{fig:si_decon_fig}. The unrelaxed surface and bulk crystal diamond were part of the database and accurately fitted, as well as some configurations along the path. The ACE energy is significantly smoother than GAP and closer to DFT reference.

\subsubsection{Point defects}
The diamond vacancy formation energy and interstitial formation energies including tetragonal, hexagonal and dumbbell are shown in Table.~\ref{tbl:si_table}. Both ACE and GAP predict point defect formation energies very well.

\paragraph{Fourfold defect}

\begin{figure}
    \includegraphics[width=\columnwidth]{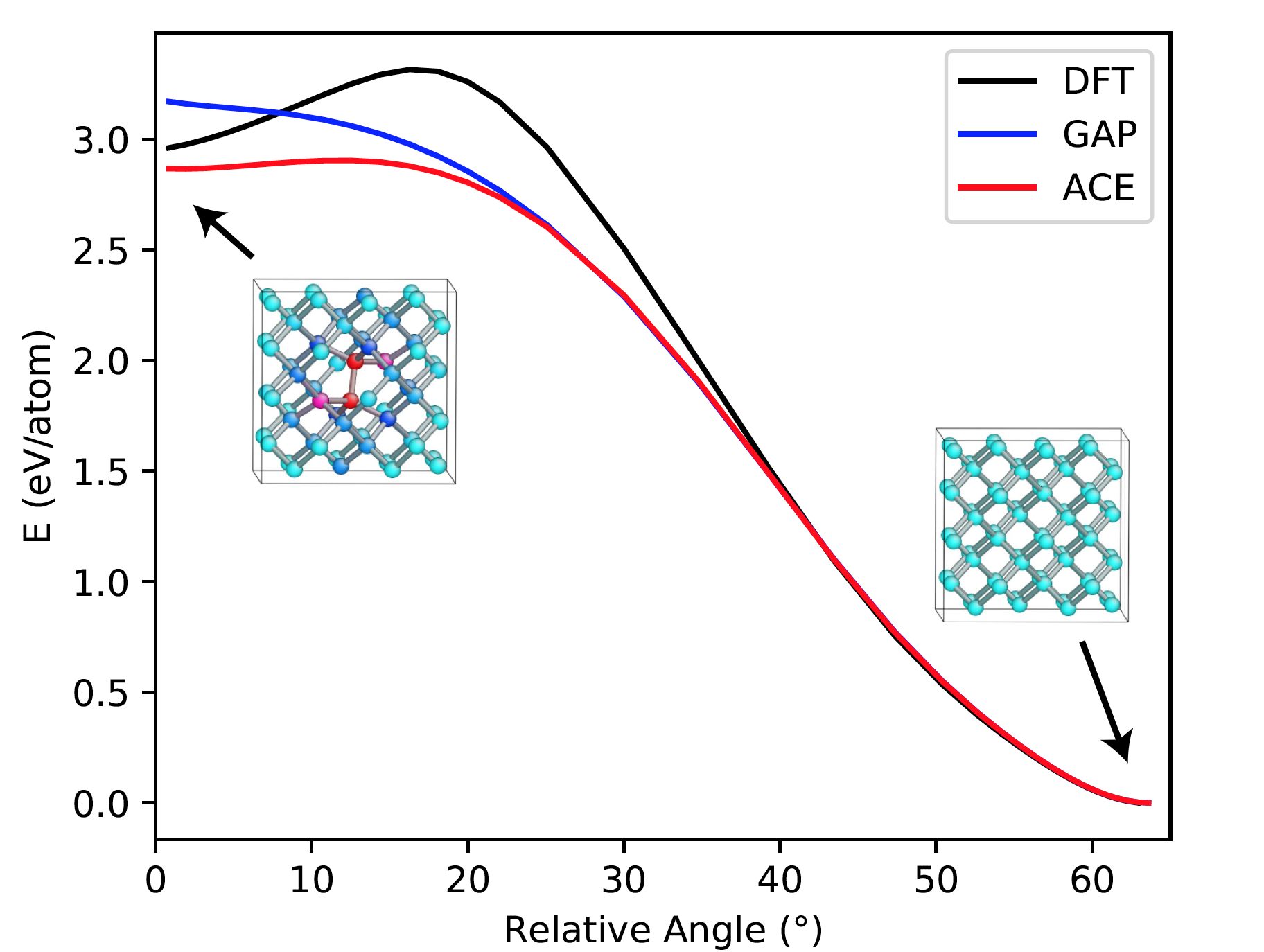}
    \caption{{\bf Fourfold-coordinated defect.} Minimum energy path connecting the fourfold defect to bulk crystal silicon evaluated using the ACE model in a 64 atom cell. We also show the energies of GAP and DFT on the exact same path. 
    }
    \label{fig:si_fourfold_defect}
\end{figure}

The lowest formation energy point defect is the ``fourfold-coordinated defect'' which consists of a bond rotation followed by a reconnection of all broken bonds\cite{Goedecker02}.  We performed the following test using the ACE model: optimise the defect structure (using a 64 atom cell) with DFT, then re-optimise it with ACE, and finally compute the minimum energy transition path to the perfect crystal. When this test was performed with GAP in \citet{Bartok18}, no local minimum was found corresponding to the defect. With ACE however, we do find a local minimum, as shown in Fig.~\ref{fig:si_fourfold_defect}, where we also show the energy of the path evaluated with DFT and GAP. Remarkably, while both GAP and ACE make a similar error near the transition state, the ACE energy is significantly better for the relaxed defect structure, leading to the stabilisation of the defect. Note that there are no configurations in the fitting database (which is identical for GAP and ACE) near the defect structure and the transition state, so again this shows the extrapolation power of the ACE model. 

\begin{figure}
    \includegraphics[width=\columnwidth]{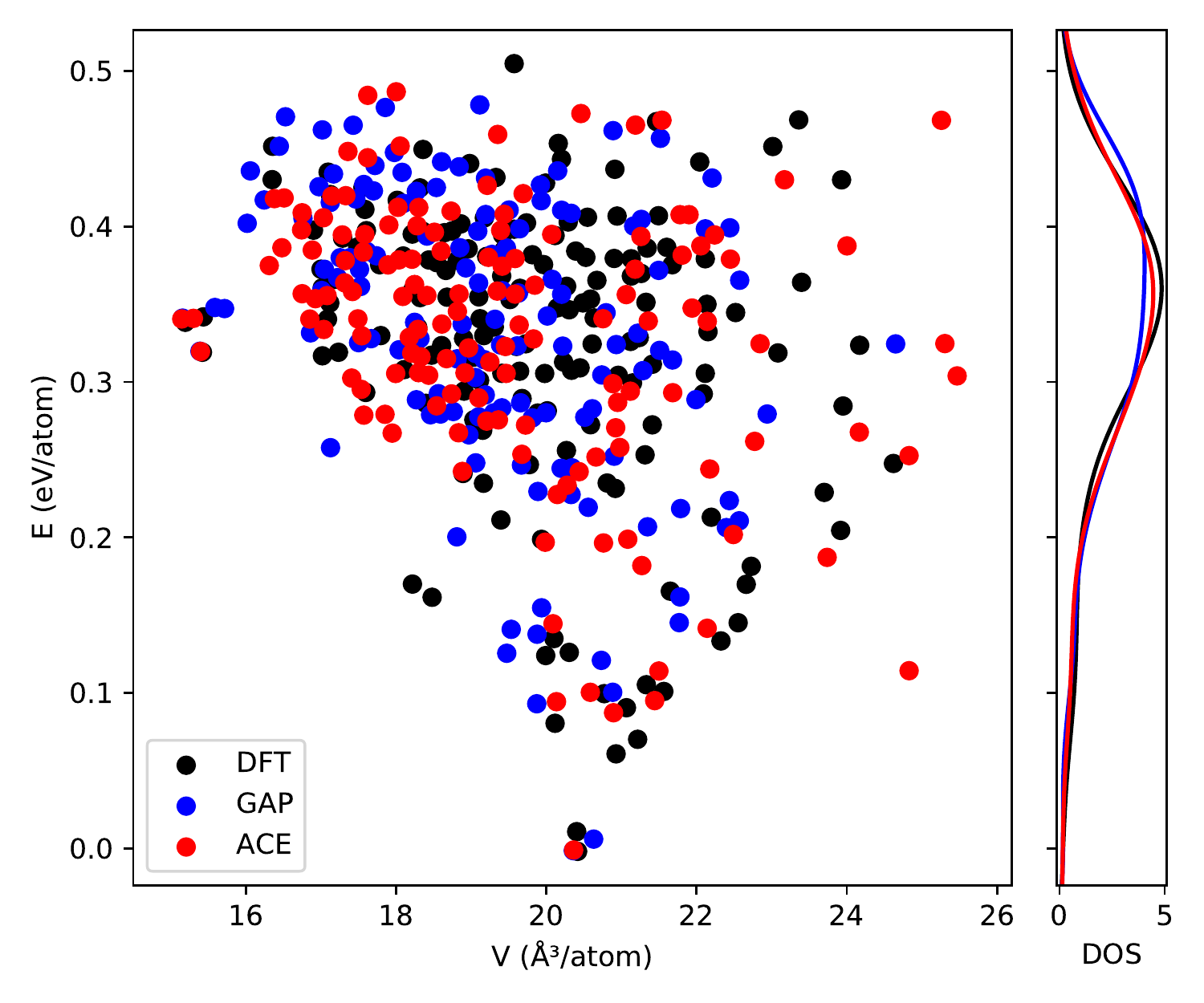}
    \caption{{\bf Random structure search.} Relaxed volumes and energies relative to diamond. The right panel shows the density of states.}
    \label{fig:si_rss_fig}
\end{figure}

\paragraph{Random structure search}
In the random structure search (RSS) \cite{Pickard06,Pickard11} test, randomized atomic configurations are relaxed, providing a view of the fitted potential energy surface for higher energy configurations. The RSS tests were performed on eight-atom configurations with close to cubic initial shapes and initial interatomic distances $>1.7$  \AA. These configurations were then relaxed using the two-point steepest-descent method \cite{2SD}. The resulting energy per atom versus volume per atom distribution is shown in Fig.~\ref{fig:si_rss_fig}. ACE shows a similar distribution compared to DFT, with the diamond structure at the correct volume and a few structures up to 0.2 eV per atom higher at comparable or somewhat larger volumes. A larger group of configurations is found at higher energies over a wider distribution of volumes. The density of states on the right panel of Fig.~\ref{fig:si_rss_fig} shows excellent agreement with DFT, as does the GAP model. This test is a strong discriminator between potentials, with all empirical potentials tested in \citet{Bartok18} failing completely. 

\subsection{Discussion \label{sec:discussion}}

We present a performant implementation of ACE in the form of the \verb+PACE+ code. We demonstrate that ACE, as implemented in \verb+PACE+, shifts the Pareto front to higher accuracy and faster evaluation times, as compared to a number of machine learning potentials from Ref.~\onlinecite{Zuo20}. 
For our general purpose parameterizations of Cu and Si the CPU time per atomic force call is below 1 ms. 
As our implementation is fully compatible with LAMMPS, large scale simulations become possible, which we demonstrate through the computation of the free energies of liquid and solid phases for evaluating the melting temperature. 
\verb+PACE+ provides a simple interface for implementing non-linear functions {$\calF$} (Eq.~\ref{eq:generalF}) as well as arbitrary radial functions which enables to adapt quickly to future ACE parameterizations.

We choose two distinct elements to illustrate parameterizations of ACE. Copper, for which classical potentials such as EAM are known to provide a good description of the interatomic interaction, and Si. For Si many different potentials were published to date and a recent GAP was shown to perform significantly better than other potentials\cite{Bartok18}. We compare our Cu parameterization to a very good EAM potential and recent GTINV and SNAP potentials. The ACE for Si is compared to GAP.

For copper, EAM provides a very good description of the energy and ACE improves on this in particular for bonding environments that require angular contributions. Excellent extrapolation to new atomic environments is demonstrated, for example, for the phonons in a free standing Cu monolayer in Fig.~\ref{fig:Cu_2D_hex_phonon_dispersion_DOS_supercell5}. Furthermore, the longer cutoff of ACE enables us to reproduce bond breaking and making in accurate agreement with the DFT reference data. It appears that SNAP and GTINV were parameterized to selected reference data, which leads to deviations from DFT in several of our tests. 

The Si ACE is comparable in accuracy to GAP, with a few key improvements. The ACE hypersurface is smoother than GAP, which is important in particular for extrapolation to large volumes as shown in Fig.~\ref{fig:si_extrap_bulks_fig}. The improved smoothness can also be seen in the surface decohesion curve in Fig.~\ref{fig:si_decon_fig} showing behavior closely matching the DFT reference. Another example of the ACE extrapolation is the fourfold defect which was highlighted in the original GAP paper, predicted erroneously to be unstable. However, ACE was fitted to the exact same DFT database and does predict a stable fourfold defect. Furthermore, it is notable that this Si ACE potential is approximately 30 times faster than the GAP of \citet{Bartok18}.

\section{Methods}

We give detailed expressions for energies and forces and efficient algorithms for their evaluation in \verb+PACE+ in the following. For Cu and Si we employ distinct ACE forms, and different parameterization strategies follow for the two elements. The details of the parameterization strategy are provided in the SI.

\subsection{Expressions for energy and forces \label{sec:energyforces}}
\subsubsection{Energy}
The energy of atom $i$ is given by
\begin{equation}
E_i = \calF(\ace_i^{(1)}, \dots,\ace_i^{(P)} ) \,,\label{eq:EF}
\end{equation}
where $\calF$ is a general non-linear function that may be supplied. Each atomic property $\ace^{(p)}_i$ is given by an ACE expansion, which is obtained as follows: given the relative neighbor positions $\pmb{r}_{ji} = \pmb{r}_{j} - \pmb{r}_{i}$,  $r_{ji} = | \pmb{r}_{ji} |$ and directions $\hat{\pmb{r}}_{ji} = \pmb{r}_{ji}/r_{ji}$, we first evaluate the atomic base
\begin{equation}
    A_{i \mu n l m} = \sum_j \delta_{\mu \mu_j} \phi_{\mu_j\mu_i nlm}(\pmb{r}_{ji}) \,,\label{eq:A}
\end{equation}
where the one-particle basis $\phib$ is given in terms of spherical harmonics $Y_{lm}(\hat{\pmb{r}}_{ji}) $ and radial functions $R_{nl}^{\mu_j\mu_i} (r_{ji})$ by
\begin{equation}
\phi_{\mu_j \mu_i nlm} = R_{nl}^{\mu_j \mu_i}(r_{ji}) Y_{lm}(\hat{\pmb{r}}_{ji}) \,. \label{eq:ab}
\end{equation}
Permutation-invariant many-body basis functions are obtained by forming products, 
\begin{equation}
    {\AAb}_{i\pmb{\mu n l m}} = 
    \prod_{t = 1}^{\ord} A_{i \mu_t n_t l_t m_t} \,. \label{eq:B}
\end{equation}
The body order of the products is $\ord+1$ and the species of atom $i$ is $\mu_i$. The vectors $\pmb{\mu}$, $\pmb{n}$, $\pmb{l}$ and $\pmb{m}$ have length $\ord$ and contain atomic species, radial function indices, and spherical harmonics indices, respectively. The ACE expansion of an atomic property $\ace_i^{(p)}$ is now given by 
\begin{equation}
    \ace_i^{(p)} =     \sum_{\pmb{\mu n l m}} \cAA^{(p)}_{\mu_i\pmb{\mu n l m}} {\AAb}_{i\pmb{\mu n l m}}  \label{eq:rho},
\end{equation}
with expansion coefficients $\cAA^{(p)}_{\mu_i \pmb{\mu n l m}}$ and lexicographically ordered indices $\pmb{\mu n l m}$.

The coefficients $\cAA^{(p)}_{\mu \pmb{\mu n l m}}$ are {\em not} free model parameters to be fitted since the $\AAb$ basis does not satisfy all required symmetries. An isometry invariant basis $\BBb$ is obtained by coupling elements of the $\AAb$ basis through the generalized Clebsch--Gordan coefficients, $\BBb = \mathcal{C} \AAb$, which yields a linear model 
\[
    \ace_i = \cBB^T \BBb = \cBB^T \mathcal{C} \AAb \overset{!}{=} \cAA^T \AAb,
\]
from which we obtain $\cAA = \mathcal{C}^T \cBB$. The $\cBB$ coefficients are the free model parameters that are optimised in the fit. We refer to \citet{Drautz19,Drautz20,Dusson20} for details. 
It is helpful to think of the expansion coefficients $\cAA^{(p)}_{\mu_i \pmb{\mu n l m}}$ as satisfying linear constraints that ensure invariance of the properties $\ace_i$ and hence of the energy under rotation and inversion.

\subsubsection{Forces}
The force on atom $k$ is written as
\begin{equation}
    \pmb{F}_k = \sum_i \left( \pmb{f}_{ik} - \pmb{f}_{ki} \right), 
\end{equation}
and the pairwise forces $\pmb{f}_{ki}$ obtained using an adjoint method, 
\cite{Drautz19,Drautz20}
\begin{align}
    \notag 
    \pmb{f}_{ki} &:= \nabla_{\br_{ki}} E_i  \\ 
    \label{eq:fkiA}
    &= \sum_{\mu n l m} \omega_{i\mu n l m} \nabla_{\br_{ki}} {A}_{i \mu n l m}  \\ 
    \label{eq:fki}
    &= \sum_{nlm} \omega_{i \mu_k nlm} \nabla_k \phi_{\mu_k \mu_i n l m}(\br_{ki}) \,,
\end{align}
where the adjoints $\omega_{i\mu nlm}$ are given by 
\begin{align}
    {\omega}_{i\mu n l m}
    &= \sum_{\pmb{\mu n l m}} \Theta_{i\pmb{\mu n l m}}  \sum_t d\AAb_{i\pmb{\mu n l m}}^{(t)} \,,
    \label{eq:omega} \\ 
d\AAb_{i\pmb{\mu n l m}}^{(t)} 
&= \delta_{\mu  \mu_t}  \delta_{n n_t}  \delta_{l l_t}  \delta_{m m_t}  %\nonumber \\ 
\prod_{s \neq t}  A_{i \mu_s n_s l_s m_s}\,, \label{eq:dB} \\ 
\Theta_{i \pmb{\mu n l m}} &=   \sum_p \pder{\calF}{\ace_i^{(p)}} \cAA^{(p)}_{\mu_i \pmb{\mu n l m}}  \,. \label{eq:Xi}
\end{align}

\subsubsection{Additional Symmetries}
A straightforward opportunity for optimisation arises due to the fact that the product basis functions fulfill
\begin{equation} \label{eq:eval_real_optim}
\operatorname{Re}\left(\AAb_{\pmb{\mu n l m}} \right) = (-1)^{\sum_t {m_t}}\operatorname{Re}\left(\AAb_{\pmb{\mu n l} -\pmb{m}} \right),
\end{equation}
and $\sum_t {m_t}=0$ for rotational invariance. As we are interested in a real-valued expansion, this identity is exploited by combining the $\AAb_{\pmb{\mu n l m}}$ and $\AAb_{\pmb{\mu n l} -\pmb{m}}$ and thus reducing the computational effort for evaluating the product basis by nearly 50\%.

Similarly, when evaluating the forces only the real part needs to be evaluated, as the imaginary part has to add up to zero. Since
\begin{equation}
\nabla {\phi}_{ \mu_j \mu_i n l -m}(\pmb{r}_{ji}) = (-1)^m \left(  \nabla {\phi}_{ \mu_j \mu_i n l m}(\pmb{r}_{ji}) \right)^* \,,
\end{equation}
and therefore
\begin{equation}
{\omega}_{i\mu_k n l -m}   = (-1)^m \left( {\omega}_{i\mu_k n l m} \right)^* \,,
\end{equation}
one can limit the force evaluation to
\begin{align}
\pmb{f}_{ki} &=  \sum_{n l, m=0} \operatorname{Re}( {\omega}_{i\mu_k n l 0} ) \operatorname{Re}(  \nabla_k {\phi}_{ \mu_k \mu_i l 0}(\pmb{r}_{ki}) ) \nonumber \\
&+ 2 \sum_{n l, m>0}  \operatorname{Re}( {\omega}_{i\mu_k n l m}  \nabla_k {\phi}_{ \mu_k \mu_i l m}(\pmb{r}_{ki}) ) \,, \label{eq:fki2}
\end{align}
which saves about 75\% of the multiplications compared to fully evaluating all complex terms.

\subsection{Algorithms \label{sec:workflow}}

\subsubsection{Model specification ({\tt PACE} Input Parameters)}
A \verb+PACE+ model is specified through four ingredients:
\begin{enumerate}
    \item specification of the radial basis, typically as splines or through a polynomial recursion
    \item a list of basis functions identified through $\pmb{\mu n l m}$ for each required order $\ord$
    \item the corresponding expansion coefficients $\cAA^{(p)}_{\mu_i \pmb{\mu n l m}}$
    \item The nonlinearity $\calF(\ace_i^{(1)}, \dots,\ace_i^{(P)} )$ and its derivatives ${\partial\calF}/{\partial\ace^{(p)}}$
\end{enumerate}

\subsubsection{Compressed basis representation\label{sec:memory}}
To formulate the evaluation algorithms it is convenient to reorganize the basis specification into a ``compressed'' format. First, we enumerate the list of one-particle basis functions and the atomic base $A$ by identifying 
\[
    \vi \equiv (\mu, n, l, m), \qquad \text{and} \qquad 
    A_{i\vi} \equiv A_{i\mu n lm}.
\]
A tuple $\vii = (\vi_1, \dots, \vi_\ord)$ can then be identified with 
${\bm \mu} {\bm n} {\bm l} {\bm m}$ and specifies a corresponding many-body basis function
\[
    \AAb_{i\vii} = \prod_{\alpha = 1}^r A_{i \vi_\alpha} 
        \equiv \AAb_{i {\bm \mu}{\bm n}{\bm l}{\bm m}} \,.
\]
An atomic property $\ace_i$ can now be written more succinctly as 
\[
    \ace^{(p)}_i = \sum_{\vii} \cAA^{(p)}_{\mu_i\vii} \AAb_{i\vii}.
\]
This format condenses the notation as well as simplifies the basis specification, now given by (i) a list of one-particle basis functions indexed by $\vi$; and (ii) a list of many-body basis functions, each represented by a tuple $\vii = (\vi_1, \dots, \vi_\ord)$ where the length $\ord$ specifies the interaction order.

\subsubsection{Force and energy evaluation \label{sec:fe}: Summary}

The energy and force for a given atom $i$ are obtained in five steps:
\begin{enumerate}
    \item Evaluate atomic base $A_{i \mu n l m}$, Eq.(\ref{eq:A}).
    \item Evaluate product basis $\AAb_{i \vii}$,
    Eq.(\ref{eq:B}) \\ 
    and properties $\ace_i^{(p)}$, Eq.(\ref{eq:rho}).
    \item Obtain energy $E_i$, Eq.(\ref{eq:EF}), and its derivatives with respect to the properties {$\ace_i^{(p)}$}.
    \item Compute product basis function derivatives $d\AAb^{(t)}_{i\vii}$, Eq.(\ref{eq:dB}), and adjoints ${\omega}_{i\mu n l m}$, Eq.(\ref{eq:omega}).
    \item Assemble forces $\pmb{f}_{ji}$, Eq.(\ref{eq:fki}).
\end{enumerate}
In the following we summarize algorithms for an efficient implementation.

\subsubsection{Optimisations}
Although we will not go into details of performance oriented code optimisations, we briefly mention the four most important ingredients: (1) recursive algorithms to evaluate the polynomial, radial and spherical basis sets; (2) contiguous memory layout for the many-body basis specification; (3) recursive evaluation of the many-body basis (cf. \S~\ref{sec:recursiveeval}); and (4) reducing the basis size and force evaluation by exploiting that the expansions are real; cf. Eqs.~\eqref{eq:eval_real_optim} and \eqref{eq:fki2}. 

\subsubsection{Stage 1: Atomic base $A$}

First the radial functions, spherical harmonics and their respective gradients are obtained. The spherical harmonics are computed in cartesian coordinates directly \cite{Drautz20}. Then the atomic base is evaluated. For $-l \leq m \leq 0$ we exploit
\begin{equation}
A_{i \mu n l -m} = (-1)^m A_{i \mu n l m}^* \,. 
\end{equation}
Note that only for evaluations of the atomic base we need to work in $\mu n l m$ notation.

\begin{algorithm}[H]
\caption{\label{algo:A} Atomic base $A$}
    \begin{algorithmic}[l]
        \State $A_{i \mu n l m}=0$
        \For{$j \leftarrow$ neighbors of atom $i$}
             \State $\mu$ = type of atom $j$
             \State compute $R_{nl}$, $dR_{nl}$, $Y_{lm}$, $dY_{lm}$
             \For{ $n, l, (m \ge 0)$ }
                 \State $A_{i \mu n l m}\mathrel{{+}{=}}R_{n l} \cdot Y_{l m}$
             \EndFor
        \EndFor         
        \For{ n, l, (m $>$ 0) }
                 \State $A_{i \mu n l -m} =  (-1)^m A_{i \mu n l m}^{*}$
        \EndFor
    \end{algorithmic}
\end{algorithm}
For numerical efficiency all pairwise radial functions can be represented as splines with several thousands interpolation points, which makes it possible to implement arbitrary radial basis functions efficiently.

\subsubsection{Product basis functions $\AAb$, expansions $\ace$ and energy}

Next the product basis functions $\AAb$ and their derivatives Eq.~(\ref{eq:dB}) are set up. 

\begin{algorithm}[H]
    \caption{\label{algo:AA} Product basis functions $\AAb$}
    \begin{algorithmic}[1]
        \State $\ace_i^{(p)} = 0$
        
        \For{$\vii \leftarrow $\pmb{basis}, $\ord = {\rm len}(\vii)$}
            \State $\AAb_{i\vii} = {\rm Re}(\prod_{t = 1}^{\ord} A_{i \vi_t})$
            \For{$p$}
                \State $\ace_i^{(p)} +\!\!= \cAA^{(p)}_{\mu_i \vii}\cdot \AAb_{i\vii}$
            \EndFor
        \EndFor
    \end{algorithmic} 
\end{algorithm}

The atomic properties $\ace^{(p)}_i$ are computed following Eq.(\ref{eq:rho}). Because $\AAb_{i\vii}$ is used only to construct the $\ace^{(p)}_i$, it need not be stored. Next, the energy $E_i = \calF(\ace_i^{(1)}, \dots,\ace_i^{(P)})$ and its derivatives $\partial \calF/\partial \ace^{(p)}_i$ are obtained.

\subsubsection{Adjoints $\omega$}
Once the derivatives $\partial \calF/\partial \ace^{(p)}_i$ are known, the adjoints ${\omega}_{i\mu n l m}$ are computed following Eq.(\ref{eq:omega}). 

\begin{algorithm}[H]
    \caption{ \label{algo:weights} Adjoints $\omega$}
    \begin{algorithmic}[l]
    \State ${\omega}_{i k} = 0$

    \For{$\vii \leftarrow $\pmb{basis}, $\ord = {\rm len}(\vii)$}
        \State $\Theta_{i\vii} = \sum_{p} \partial \calF/\partial \ace^{(p)}_i \cdot \cAA^{(p)}_{\mu_i \vii}$
        \State compute $d\AAb_{i\vii t}$, $t = 1, \dots, \ord$ (Alg. 4)
        \For{ $t \leftarrow  1, \dots, \ord$}
            \State $\omega_{i \vi_t} +\!\!=  \Theta_{i\vii} \cdot {\rm Re}({d\AAb}_{i \vii t})$
        \EndFor  
    \EndFor
    \end{algorithmic} 
\end{algorithm}

Here, $\Theta_{i \vii}$ and $d\AAb_{i\vii t}$ are only required locally and stored in a temporary variable. The derivatives  $d\AAb_{i\vii t}$ can be computed via backward differentiation with cost that scales linearly in $\ord$ instead of the $O(\ord^2)$ scaling for a naive implementation. This important optimisation is implemented as follows: 

\begin{algorithm}[H] 
    \caption{\label{algo:dAAt} Compute $d\AAb_{i\vii t}$, $t = 1, \dots, \ord$}
    \begin{algorithmic}[1]
          \State $\textbf{A}^{\rm fwd} = 1$;
          ${d\textbf{A}}_{i \vii 1} = 1$ \;
          \For{$t \leftarrow 1, 2, \dots, \ord-1$}
              \State $\textbf{A}^{\rm fwd} *\!\!= A_{i \vi_t}$ \;
              \State ${d\textbf{A}}_{i \vii (t+1)} 
                        = \textbf{A}^{\rm fwd}$ \;
          \EndFor
          \State $\textbf{A}^{\rm bwd} = 1$ \;
          \For{$t \leftarrow \ord, \ord-1, \dots, 2$}
           \State $\textbf{A}^{\rm bwd} *\!\!= A_{i \vi_t}$
           \State ${d\textbf{A}}_{i \vii (t-1)} 
            *\!\!= \textbf{A}^{\rm bwd}$ \;
          \EndFor 
    \end{algorithmic} 
\end{algorithm}

The computation of $ d{\mathbf A}$ can be slightly improved by removing multiplications by one inside the loop. With that optimisation the number of multiplications scales as $3\ord-5$ for $\ord \geq 2$.

\subsubsection{Pairwise forces}
The gradients may now be obtained from Eq.~\eqref{eq:fki2}.   

\begin{algorithm}[H] 
    \caption{\label{alg:forces} Compute $\pmb{f}_{ki}$}
    \begin{algorithmic}[1]
        \State $\pmb{f}_{ki} = 0$
        \For{ $n, l, (m \ge 0)$ }
        \For{ $k  \leftarrow$ {neighbors of atom} $i$ }
            \State $\pmb{f}_{ki} +\!\!= (2-\delta_{m0}){\rm Re}\big({\omega}_{i\mu_k n l m}  \nabla_k {\phi}_{\mu_k \mu_i n l m}(\pmb{r}_{ki})\big)$
        \EndFor
        \EndFor
    \end{algorithmic} 
\end{algorithm}

\subsubsection{Computational cost}
\label{sec:cost_standard}
The overall computational cost is composed of two essentially independent contributions: (i) the evaluation of the atomic base $A$ requires $O(N \cdot \#A)$ evaluations; and (ii) the evaluation of the correlations $\AAb$ requires $O((\maxord + \#\ace) \cdot \#\AAb)$ evaluations. That is, the overall cost scales linearly in the number of neighbors $N$, the maximum correlation order $\maxord$ and also linearly in the number of properties $\ace_i^{(p)}$. We will see next that the $\ord$-dependence can be further reduced with an alternative evaluation scheme.

\subsection{Recursive evaluator}
\label{sec:recursiveeval}
In most cases, the evaluation of the product basis functions and their derivatives (Algorithms~\ref{algo:AA}, \ref{algo:weights} and \ref{algo:dAAt}) are the computational bottleneck. Here, we detail the implementation of an alternative {\em recursive} evaluation algorithm\cite{Dusson20}, reminiscent of dynamic programming concepts, which significantly reduces the number of arithmetic operations at the cost of introducing additional temporary storage requirements.

The idea is to express the basis functions of higher correlation order in terms of a product of just two basis functions of a lower correlation order. Consider the many-body basis described as in Section~\ref{sec:memory} in terms of $\vii$ tuples indexing into the atomic base $A$. We say that a tuple $\vii$ of length $\ord$ has a decomposition 
$\vii \equiv {\vii'} \cup \vii''$, where $\vii', \vii''$ have lengths $\ord', \ord''$, if the {{tuples}} $(\vi_1, \dots, \vi_\ord)$ and $(\vi_\ord', \dots, \vi_{\ord'}', \vi_1'', \dots, \vi_{\ord''}'')$ agree up to permutations. In this case, we can write 
\begin{equation} \label{eq:recursion_idea}
    \AAb_{i\vii} = \AAb_{i\vii'}  \AAb_{i\vii''};
\end{equation}
that is, the basis function $\AAb_{i\vii}$ can be computed with a single product instead of $\ord-1$ products, while its adjoint requires no additional products.

\subsubsection{Graph construction}
A key subtlety must be addressed before putting this into practise: Due to the constraints that $\sum_t m_t = 0$ and $\sum_t l_t$ even, not all basis functions have a decomposition  \eqref{eq:recursion_idea} that respects those constraints. For example, if ${\bm m} = (1,0,-1)$ then we may decompose it as $(1,-1) \cup (0,)$, but if ${\bm m} = (2,-1,-1)$ then no such decomposition exists. To overcome this we add ``artificial'' basis functions to the model supplied with zero coefficients. A simple but seemingly effective heuristic how to achieve this efficiently is described in the following. The result of this construction is a directed acyclic graph
\[
    \mathcal{G} = \big\{  \vii \equiv \vii' \cup \vii'' \big\} \,,
\] 
where each node $\vii$ represents a basis function supplied with coefficient $\tilde{\bm c}^{(p)}_{\mu_i \vii}$ with exactly two incoming edges $(\vii', \vii), (\vii'', \vii)$ and arbitrarily many (possibly zero) outgoing edges. The values of the coefficients are readily obtained from the canonical basis representation.

To construct $\mathcal{G}$ we first insert the atomic base {\{$A_{i\vi}$\}} represented by its indices {$\{ \vi \}$}  into the graph as root nodes. Then, with increasing correlation order we insert the nodes $\vii$ using the following recursive algorithm: 

\begin{algorithm}[H]
    \label{algo:dagconstruction}
    \begin{algorithmic}[1]
    \caption{Insert node {$\vii$}  into graph $\mathcal{G}$}
    \For{$\text{all decompositions~} \vii \equiv \vii' \cup \vii''$}
    \If{$\vii', \vii'' \in \mathcal{G}$}
    \State $\mathcal{G} \leftarrow \mathcal{G} \cup \{ \vii \equiv \vii' \cup \vii'' \}$
    \State return
    \EndIf
    \EndFor
    \State identify $\vii = (\vi_1,) \cup \vii'$ where $\vi_1$ is maximal
    \State recursively insert $\vii'$ into $\mathcal{G}$
        with zero-coefficients
    \State recursively insert $\vii$ into $\mathcal{G}$
    \end{algorithmic}
\end{algorithm}

This simple heuristic already leads to excellent performance, as we report at the end of this section, but further optimisations may be possible to the graph aiming to minimize the number of {\em artificial nodes} inserted into the graph and limiting the additionally required memory access.

\subsubsection{Recursive evaluation}
To evaluate the properties $\ace_i^{(p)}$ we first apply Algorithm~\ref{algo:A} to obtain the atomic base $A_{i\vi} \equiv A_{i\mu nlm}$. Next, we can traverse the graph taking care to only evaluate basis functions whose parents have already been evaluated (this is implicitly assumed), e.g., by looping with increasing correlation order.

\newcounter{recalg}
\setcounter{recalg}{2}
\renewcommand\thealgorithm{\arabic{recalg}R}

\begin{algorithm}[H]
\label{algo:AA-rec}
    \caption{Recursive evaluation of properties}
    \begin{algorithmic}[1]
        \State $\ace_i^{(p)} = 0$
        \For{$\vii \equiv \vii' \cup \vii'' \leftarrow \mathcal{G}$}
            \State $\AAb_{i\vii} =\AAb_{i\vii'} \AAb_{i\vii''}$
            \For{$p$}
                \State $\ace_i^{(p)} \,\,+\!\!=\, \cAA_{\mu_i \vii}^{(p)} {\rm Re} (\AAb_{i\vii})$
            \EndFor
        \EndFor
    \end{algorithmic} 
\end{algorithm}

Only the values $\AAb_{i\vii}$ corresponding to {\em interior} nodes $\vii$ {(i.e. nodes that have at least one child)}, must be stored, while those corresponding to {\em leaf} nodes {(i.e. nodes without any child)} are only required locally to update $\ace_i^{(p)}$.

To evaluate the adjoints $\omega_{i \vi} \equiv \omega_{i\mu nlm}$ we use the observation that
\[
    \partial \AAb_{i \vii}
    = 
    \partial \big(\AAb_{i \vii'} \AAb_{i \vii''}\big)
    = 
    \AAb_{i \vii''} \partial \AAb_{i \vii'} 
    + 
    \AAb_{i \vii'} \partial \AAb_{i \vii''},
\]
where $\partial$ is a differential operator,
and hence
\begin{align*}
        \frac{\partial \calF(\ace_i^{(1)}, \dots,\ace_i^{(P)} )}{\partial \AAb_{i \vii}} 
        \partial \AAb_{i \vii} 
        %\sout{\sum_p {\textstyle \frac{\partial F}{\partial \ace_i^{(p)}}} \cAA_{i \vii}^{(p)}  %\AAb_{i \vii}} 
    &= 
    \overset{=: \theta_\vii}{
        \overbrace{
        \sum_p {\textstyle \frac{\partial\calF}{\partial \ace_i^{(p)}}} \cAA_{\mu_i \vii}^{(p)} }}
         \partial  \AAb_{i \vii}  \\ 
    & \hspace{-1cm} = 
    \big(\theta_{\vii} \AAb_{i \vii''} \big) \partial \AAb_{i \vii'}
    + 
    \big(\theta_{\vii} \AAb_{i \vii'} \big) \partial \AAb_{i \vii''}.
\end{align*}
This shows that the adjoint of $\AAb_{i \vii}$ can be propagated to the adjoints of the two parents. This immediately leads to the following reverse mode differentiation algorithm, which computes adjoints $\omega_{i\vii}$ for {\em all} basis functions $\AAb_{i\vii}$ (or at least those corresponding to interior nodes of the graph). However, only the adjoints for root nodes, $\omega_{i \vi} = \omega_{i \mu nlm}$, are eventually used to assemble the forces (Eq.(\ref{eq:fkiA})). The traversal of the graph must now be done in reverse order, that is, a node $\vii$ may only be visited once all of its children have been visited, for example, by traversing in reverse correlation order.

\setcounter{recalg}{3}

\begin{algorithm}[H]
\label{algo:weight-rec}
    \caption{Recursive evaluation of adjoints $\omega$}
    \begin{algorithmic}[1]
        \State $\omega_{i\vii} = 0$
        \For{$\vii \equiv \vii' \cup \vii'' \leftarrow \mathcal{G}$, in reverse order}
        \State $\theta_\vii = \sum_{p} \partial \calF/\partial \ace^{(p)}_i \cAA^{(p)}_{\mu_i \vii}$
        \State $\omega_{i\vii}\,\, +\!\!=\, \theta_\vii$
%            \For{$p$}
%                \State $\omega_{i\vii}\,\, +\!\!=\, {\textstyle \frac{\partial calF}{\partial \ace_i^{(p)}}} \cAA_\vii^{(p)}$
%            \EndFor
            \State $\omega_{i\vii'}\,\, +\!\!=\, \omega_{i\vii} \AAb_{i \vii''}$
            \State $\omega_{i\vii''} \,\,+\!\!=\, \omega_{i\vii} \AAb_{i \vii'}$        
        \EndFor
    \end{algorithmic} 
\end{algorithm}

To conclude, we now use Algorithm~\ref{alg:forces} to evaluate the forces.

\subsubsection{Computational cost of the recursive evaluator}
\label{sec:cost_recursive}
The forward pass, Algorithm~2R, requires $1+P$ multiplications and $5 + P$ memory access operations at each iteration. The backward pass, Algorithm~3R, requires $2+P$ multiplications and $7+P$ memory access operations. In particular, the cost is (seemingly) independent of the correlation order of each basis functions. The overall cost is given by 
\[
    O(N \cdot \# A) + O(\# \mathcal{G} \cdot P)\,,
\]
i.e., it scales linearly in the number of neighbors $N$ and the number of nodes in the graph. The first part for setting up the atomic base is unaffected by the recursive evaluator. 

Comparing the cost between the two approaches is difficult since we have no estimates on the number of artificial nodes that must be inserted into the graph. In practise we observe that there are always more leaf nodes than interior nodes, which means that relatively few artificial nodes are inserted and hence the recursive algorithm is significantly faster for large basis sets and high correlation order, but roughly comparable for small basis sets and low correlation order.

For the Cu potential with a smaller number of basis functions the timing decreases from 0.43 to 0.32 ms/atom/MD-step when the recursive evaluator is used. The Si potential, employing more basis functions, has a more pronounced speed-up from 1.84 ms/atom/MD-step to 0.80 ms/atom/MD-step.

\section{Data availability}

The reference data for Cu is available from \mbox{atomistictools.org/downloads}, the data for Si from libatoms.org.

\section{Code availability}

PACE is made available with the LAMMPS distribution.

\section{Author contributions}

YL, CO and RD led the implementation of PACE. YL and RD carried out the DFT calculations for Cu. The parameterization and testing for Cu was largely done by YL, MM, AB, SM and RD, for Si by CvdO, CO and GC. YL and AT ensured compatibility with LAMMPS. RD wrote the initial version of the manuscript and figures were generated by YL and CvdO. All authors contributed to the implementation of PACE, the parameterization and testing of the ACE for Cu and Si and to editing and discussing the manuscript.

\section{Competing interests}

The authors declare no competing interests.

\section{Corresponding author}

Correspondence should be addressed to Ralf Drautz.

\begin{acknowledgments}
The authors acknowledge helpful discussions with Marc Cawkwell. RD acknowledges funding through the German Science Foundation (DFG), project number 405621217.  Sandia National Laboratories is a multimission laboratory managed and operated by National Technology \& Engineering Solutions of Sandia, LLC, a wholly owned subsidiary of Honeywell International Inc., for the U.S. Department of Energy’s National Nuclear Security Administration under contract DE-NA0003525.  
\end{acknowledgments}

\bibliography{ace.bib}

\end{document}